\documentclass[useAMS,usenatbib]{mn2e}
\usepackage{graphicx}
\usepackage{xcolor}
\usepackage{multirow}
\usepackage{wasysym}  
\usepackage{sidecap}

\title[Periods and Disc structure of four AM\,CVn systems]{Orbital periods and Accretion disc structure of four AM\,CVn systems}
\author[T. Kupfer et al.]{T. Kupfer$^{1}$\thanks{E-mail:t.kupfer@astro.ru.nl}, P.~J. Groot$^{1,2}$, D. Levitan$^{2}$, D. Steeghs$^{3}$, T.~R. Marsh$^{3}$, \newauthor{R.~G.~M. Rutten$^{4}$ and G. Nelemans$^{1,5}$}\\
$^{1}$Department of Astrophysics/IMAPP, Radboud University Nijmegen, P.O. Box 9010, 6500 GL Nijmegen, The Netherlands\\
$^{2}$Division of Physics, Mathematics, and Astronomy, California Institute of Technology, Pasadena, CA 91125, USA\\
$^{3}$Department of Physics, University of Warwick, Coventry CV4 7AL, UK\\
$^{4}$GRANTECAN, Spain\\
$^{5}$Institute for Astronomy, KU Leuven, Celestijnenlaan 200D, 3001 Leuven, Belgium}
\begin{document}

\date{Accepted 2013 March 21 Received 2013 March 21; in original form 2013 January 25}

\pagerange{\pageref{firstpage}--\pageref{lastpage}} \pubyear{2002}

\maketitle

\label{firstpage}

\begin{abstract}
Phase-resolved spectroscopy of four AM\,CVn systems obtained with the William Herschel Telescope and the Gran Telescopio de Canarias (GTC) is presented. SDSS\,J120841.96+355025.2 was found to have an orbital period of 52.96$\pm$0.40\,min and shows the presence of a second bright spot in the accretion disc. The average spectrum contains strong Mg\,{\sc i} and Si\,{\sc i/ii} absorption lines most likely originating in the atmosphere of the accreting white dwarf. SDSS\,J012940.05+384210.4 has an orbital period of 37.555$\pm$0.003 min. The average spectrum shows the Stark broadened absorption lines of the DB white dwarf accretor. The orbital period is close to the previously reported superhump period of 37.9\,min. Combined, this results in a period excess $\epsilon$=0.0092$\pm$0.0054 and a mass ratio $q=$0.031$\pm$0.018. SDSS\,J164228.06+193410.0 displays an orbital period of 54.20$\pm$1.60\,min with an alias at 56.35\,min. The average spectrum also shows strong Mg\,{\sc i} absorption lines, similar to SDSS\,J120841.96+355025.2. SDSS\,J152509.57+360054.50 displays an period of 44.32$\pm$0.18\,min. The overall shape of the average spectrum is more indicative of shorter period systems in the 20-35 minute range. The accretor is still clearly visible in the pressure broadened absorption lines most likely indicating a hot donor star and/or a high mass accretor. Flux ratios for several helium lines were extracted from the Doppler tomograms for the disc and bright spot region, and compared with single-slab LTE models with variable electron densities and path lengths to estimate the disc and bright spot temperature. A good agreement between data and the model in three out of four systems was found for the disc region. All three systems show similar disc temperatures of $\sim$10\,500~K. In contrast, only weak agreement between observation and models was found for the bright spot region. 

\end{abstract}

\begin{keywords}
accretion, accretion discs -- binaries: close -- stars: individual:  -- stars:
individual: SDSS\,J120841.96+355025.2, SDSS\,J012940.05+384210.4, SDSS\,J164228.06+193410.0, SDSS\,J152509.57+360054.5.
\end{keywords}

\section{Introduction}
AM\,CVn systems are a small group of mass transferring ultra compact binaries with orbital periods between 5.4 and 65 min (see Solheim 2010\nocite{sol10} for a recent review). They consist of a white dwarf (WD) primary, and a WD, or semi-degenerate helium star secondary (e.g. Nelemans et al. 2001)\nocite{nel01}. Spectroscopically these systems are characterized by a high deficiency of hydrogen, indicating an advanced stage of binary evolution. 

Over the course of evolution of a close binary consisting of intermediate mass stars, two common envelope phases may produce a detached WD binary system at a period of minutes to hours (see \citet{kil12} and references therein). Gravitational wave radiation will decrease the orbital separation until the low-mass secondary fills its Roche lobe and mass transfer sets in at an orbital period between 3-10 min (e.g. Marsh et al. 2004). A fraction of these systems will survive the ensuing direct impact phase, pass through an orbital period minimum and become AM\,CVn systems \citep{nel01,mar04}. An accretion disc forms at an orbital period of $\sim$10 min depending on the masses and the entropy of the donor (see e.g. Kaplan et al. 2012\nocite{kap12}). Material transferred from the secondary hits the disc at the so-called bright spot. In the further evolution the system widens, upon loss of angular momentum in gravitational waves, causing the mass-transfer rate to drop. The mass-transfer rate drops as the orbit widens and the system ends up as a more massive WD with an extremely low-mass WD companion ($\sim$0.001~M$_{\odot}$) at an orbital period of $\sim$1 hour. AM\,CVn systems are important as strong low-frequency Galactic gravitational wave sources (e.g. \citet{nel04, roe07b, nis12}) and are the source population of the proposed '.Ia' supernovae \citep{bil07}.

AM\,CVn systems show three phases during their lifetime based on their photometric behaviour. Systems with periods below 20 min have a high mass-transfer rate resulting in spectra that are dominated by helium absorption lines originating in the optically thick disc. Systems with periods above 40 min are in a low mass-transfer rate state and have spectra dominated by strong helium emission lines from an optically thin accretion disc. The intermediate period systems between 20 and 40 min orbital period undergo dwarf-nova type photometric outbursts and change their spectral behaviour from low state, where emission lines from the disc and Stark-broadened absorption lines from the accretor dominate the spectrum, to a high state, where absorption lines from the optically thick disc dominate the spectra \citep{gro01,lev11,ram12}. 

In a large campaign using the Sloan Digital Sky Survey (SDSS; York et al. 2001\nocite{yor00}) \citet{roe05,roe07,and05,and08,rau10,car12} built up the population of AM\,CVn systems and significantly increased the number of known systems over the last ten years. The recent increase in the population from photometric, spectroscopic, and synoptic surveys (e.g. Levitan et al. 2011,2012,2013)\nocite{lev11,lev12,lev13} allows the derivation of population properties such as the orbital period distribution or the presence or absence of spectral features as a function of evolutionary stage. To determine the orbital period, and detect variations in the spectrum over the orbital period, phase-resolved spectroscopy is needed \citep{roe05,roe06}, although \citet{lev11} showed that the photometric period may be tied to the spectroscopic period for some systems. This is not valid for all systems. GP\,Com shows variability in the light curve which is not tied to the orbital motion \citep{mor03}. 

Here, the results of follow-up phase-resolved spectroscopy of four systems which were originally found in the SDSS spectroscopy or photometry, and confirmed by spectroscopy to be AM\,CVn systems are presented. SDSS\,J120841.96+355025.2 (hereafter SDSS\,J1208) and SDSS\,J012940.05+384210.4 (hereafter SDSS\,J0129) were found in the spectral database of SDSS due to their strong helium emission lines and lack of hydrogen \citep{and05,and08}. SDSS\,J164228.06+193410.0 (hereafter SDSS\,J1642) and SDSS\,J152509.57+360054.5 (hereafter SDSS\,J1525) were found during follow-up observations of colour selected AM\,CVn candidates from SDSS \citep{rau10}. 

As Doppler tomography \citep{mar88} remaps line intensities from the wavelength-time domain to the binary velocity frame, the technique can also be used to derive line intensities of specific regions in the binary system. \citet{ski00} used ratioed Doppler maps to study the structure of the accretion disc in the dwarf nova WZ\,Sge. This ratio-mapping has so far not been applied to Doppler tomograms to derive disc temperatures and densities in AM\,CVn type binaries. Here, for the first time, this technique is used to limit the characteristics of the accretion disc and bright spot regions under the assumption of an LTE slab model in AM\,CVn type binaries.
\begin{table*}
 \centering
 \caption{Summary of the observations of SDSS\,J1208, SDSS\,J0129, SDSS\,J1642 and SDSS\,J1525.}
  \begin{tabular}{ccccccc}
  \hline
  System & Telescope+Instrument & Exp. & Exp. time (s) &    Coverage (\AA)  & Binning  &  Resolution (\AA)\\
  \hline\hline
  {\bf SDSS\,J1208} &                       &    &     &    &  & \\
  2008/12/25  & WHT+ISIS(R300B/R316R) & 50 & 180 & 3250 - 8119 & 2x2 & 1.77 (R300B) 1.75 (R316R) \\
  2008/12/27  & WHT+ISIS(R300B/R316R) & 50 & 180 & 3250 - 8119 & 2x2 & 1.77 (R300B) 1.75 (R316R)\\
  2008/12/28  & WHT+ISIS(R300B/R316R) & 50 & 210 & 3250 - 8119 & 2x2 & 1.77 (R300B) 1.75 (R316R) \\
   \noalign{\smallskip}
  {\bf SDSS\,J0129} &                       &    &     &  &  \\
  2011/09/25  & GTC+Osiris(R1000B) & 42 & 120 & 3833 - 7877 & 2x2  & 2.15 (R1000B) \\
  2011/09/26  & GTC+Osiris(R1000B) & 40 & 120 & 3833 - 7877 & 2x2  & 2.15 (R1000B)  \\
  2011/09/27  & GTC+Osiris(R1000B) & 20 & 120 & 3833 - 7877 & 2x2  & 2.15 (R1000B)   \\
  2011/09/28  & GTC+Osiris(R1000B) & 21 & 120 & 3833 - 7877 & 2x2  & 2.15 (R1000B)   \\
   \noalign{\smallskip}
  {\bf SDSS\,J1642} &                       &    &    & \\
  2011/08/24  & GTC+Osiris(R1000B) & 32 & 180 & 3833 - 7877 & 2x2  &  2.15 (R1000B) \\
  2011/08/26  & GTC+Osiris(R1000B) & 32 & 180 & 3833 - 7877 & 2x2  & 2.15 (R1000B)  \\
  2011/08/27  & GTC+Osiris(R1000B) & 20 & 180 & 3833 - 7877 & 2x2  &  2.15 (R1000B) \\
  2012/01/25  & Hale+DoubleSpec(600/4000)  &  1 & 1200 & 3500 - 5250   & 1x1  & 2.75 (600/4000) \\
   \noalign{\smallskip}
  {\bf SDSS\,J1525} &                       &    &   &  &\\
  2011/07/24  & GTC+Osiris(R1000B) & 16 & 180  &  3833 - 7877 & 2x2 & 2.15 (R1000B) \\
  2011/08/28  & GTC+Osiris(R1000B) & 22 & 180  & 3833 - 7877 & 2x2 & 2.15 (R1000B) \\
  2011/08/29  & GTC+Osiris(R1000B) & 32 & 180 & 3833 - 7877  & 2x2 & 2.15 (R1000B) \\
   \hline
\end{tabular}
\label{observ}
\end{table*}

\section{Observations and Data reduction}
Phase-resolved spectroscopy of SDSS\,J1208 using the William Herschel Telescope (WHT) and the ISIS spectrograph \citep{car93} was obtained. Table~\ref{observ} gives an overview of all observations and the instrumental set-ups. Each night an average bias frame was made out of 20 individual bias frames and a normalised flat field frame was constructed out of 20 individual lamp flat fields. CuNeAr arc exposures were taken every hour to correct for instrumental flexure. Each exposure was wavelength calibrated by interpolating between the two closest calibration exposures. A total of about 30 lines could be well fitted in each arc exposure using a Legendre polynomial of order 4 resulting in 0.08\,\AA\ root-mean-square residuals.

Feige 34 \citep{oke90} was used as a spectrophotometric standard to flux calibrate the spectra and correct them for the instrumental response. 


For SDSS\,J0129, SDSS\,J1642 and SDSS\,J1525 phase-resolved spectroscopy using the Gran Telescopio de Canarias (GTC) and the Osiris spectrograph \citep{cep98} was obtained. 10 to 12 bias frames each night to construct an average bias frame and 5 to 7 individual tungsten lamp flat fields to construct a normalised flat field were obtained. HgArNe calibration were obtained for SDSS\,J1642 and SDSS\,J1525 at the end of every night. To check the stability of Osiris the O\,{\sc i} 5577 sky line was used. It shows a maximum shift of 0.3\AA\ over the night. This corresponds to 17\,km\,s$^{-1}$ which is acceptable for the velocity fields displayed by our systems. For SDSS\,J0129, calibrations were obtained directly following a two hour block of the science observations. About 30 lines could be well fitted in each arc exposure using a Legendre function of order 4 resulting in 0.07\,\AA\ root-mean-square residuals. The only exception was August 26 on SDSS\,J1642 when the calibration lamp spectra were stretched by a few pixels. The reason for the stretched image was an incorrect focus position of the spectrograph. This was initially not spotted, and corrected for by adjusting the telescope focus. This produced a focussed spectrum on the detector, but had the side effect of generating a few pixels of stretching. Sky lines could not be used for calibration as to few isolated sky lines are visible in the spectrum in particular in the blue. Calibration lines from a different night could also not be used as the stretching of the spectra is a non-linear effect and variable from night to night. Therefore, the helium emission lines from the target itself were used to calibrate the spectra for this night. To correct for the instrumental response G24--9, L1363--3 and G158--100  \citep{oke74,oke90} were used as spectrophotometric standards.

All spectra were debiased and flat fielded using {\sc iraf} routines. One dimensional spectra were extracted using optimal extraction and were subsequently wavelength and flux calibrated. 

SDSS\,J1642 was also observed on the Hale telescope using the DoubleSpec spectrograph \citep{oke82}, which covers wavelengths below 3900\,\AA.  Only data taken with the blue side, using the 600/4000 grating is shown here. An FeAr lamp was used for wavelength calibration and G191-B2B \citep{oke90} was used as a spectrophotometric standard. {\sc Starlink}, {\sc pamela}, and {\sc molly} were used for reduction and calibration. 

\section{Methods}
\subsection{Period determination}\label{sec:period_det}
To determine the orbital period the violet-over-red method (V/R) described in \citet{nat81} was used following \citet{roe05,roe06,roe07a}. The normalised emission wings of the helium lines were divided into a red and a violet part and the flux ratio of both wings was calculated.  To maximise the SNR the ratios of the strongest emission helium lines (3888\,\AA, 4026\,\AA, 4471\,\AA, 5015\,\AA, 5875\,\AA\ and 6678\,\AA) were summed for SDSS\,J1208, as well as for SDSS\,J1642 without 3888\,\AA. For SDSS\,J0129 and SDSS\,J1525 the lines which are supposed to be least affected by broad wing absorption from the accretor were used (5875\,\AA, 6678\,\AA\ and 7065\,\AA). Lomb-Scargle periodograms of the measured violet-over-red-ratio as a function of the heliocentric date were computed (Fig.~\ref{fig:lombscargle}) and the strongest peaks are assumed to correspond to the orbital period. 

The uncertainty on the derived period for each system was estimated using a simple Monte Carlo simulation, where 1000 periodograms were computed and in each the highest peak was taken as the orbital period (see Roelofs et al. 2006)\nocite{roe06}. For SDSS\,J1208, 150 spectra were randomly picked out of the full sample of 150 spectra, allowing for a spectrum to be picked more than once. The same was done for SDSS\,J1029, SDSS\,J1642 and SDSS\,J1525 using 123, 84 and 70 spectra respectively. The standard deviation on the distribution of the computed orbital periods is taken as a measure of the accuracy in the derived period.

\begin{figure*}
\begin{center}
\includegraphics[width=0.95\textwidth]{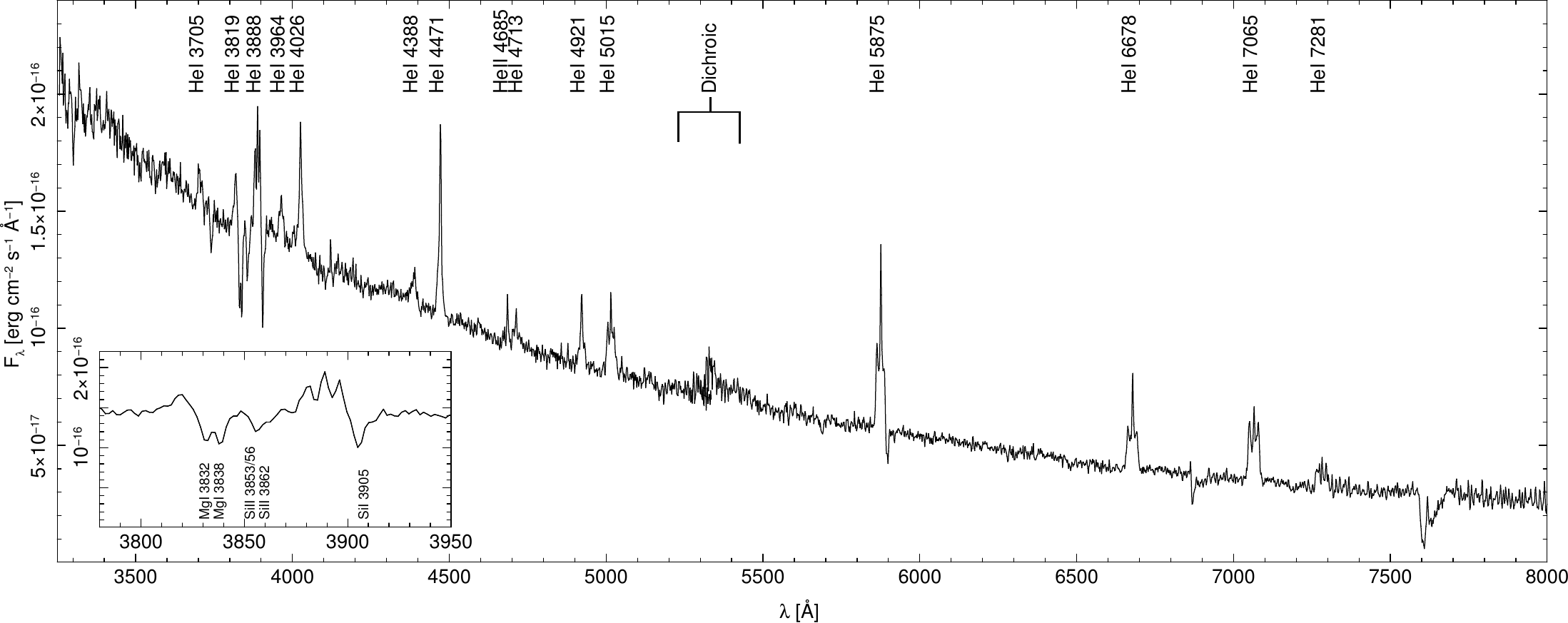}
\caption{Average spectrum of SDSS\,J1208 obtained with the WHT. The strong absorption lines of Mg\,{\sc i} and Si\,{\sc i/ii} are prominent}
\label{whtaver}
\end{center}  
\end{figure*}
\subsection{Analysis of Doppler tomograms}\label{sec:doppler_anal}
In Doppler tomography \citep{mar88} phase-resolved spectra are projected onto a two-dimensional map in velocity coordinates. We refer to \citet{ste03} and \citet{mar01} for reviews on Doppler tomography. Emission features that are stationary in the binary frame add up constructively in a Doppler tomogram while emission that is not stationary in the binary frame or moves on a period different from the orbital period will be spread out over the Doppler tomogram. Therefore, Doppler tomograms are useful to separate out features that move with a different velocity and/or different phase (e.g. bright spot and central spike). In this analysis Doppler tomograms  were computed using the software package {\sc Doppler}\footnote{{\sc Doppler} was written by Marsh and is available at http://www.warwick.ac.uk/go/trmarsh/software/}, and were used to measure the phase shift between the two bright spots in SDSS\,J1208 (see Sec.~\ref{sec:2ndbrightspot}), and to extract fluxes from Doppler tomograms to estimate temperatures of the disc and the bright spot. In the analysis we assume that the disc emission is radially and azimuthally symmetric around the WD and the bright spot emission is concentrated in a single quarter in the Doppler tomogram.

SDSS\,J1208 shows a prominent second bright spot in several helium lines (see Fig.~\ref{fig:1208dopplermap}). To measure the phase shift between both spots in SDSS\,J1208, a 2D-Gaussian was fitted to the spots in the Doppler tomograms of the 5875\,\AA, 6678\,\AA\ and 7065\,\AA\ lines. The center of every 2D-Gaussian fit ($v_{x,max},v_{y,max}$) was calculated for each line. The error-weighted mean of the center is taken to be the center of the bright spot. The phase error on the position of the bright spot is calculated as the length of the intersection between the phase angle and the positional error ellipse.

\begin{figure}
\begin{center}
\includegraphics[width=0.49\textwidth]{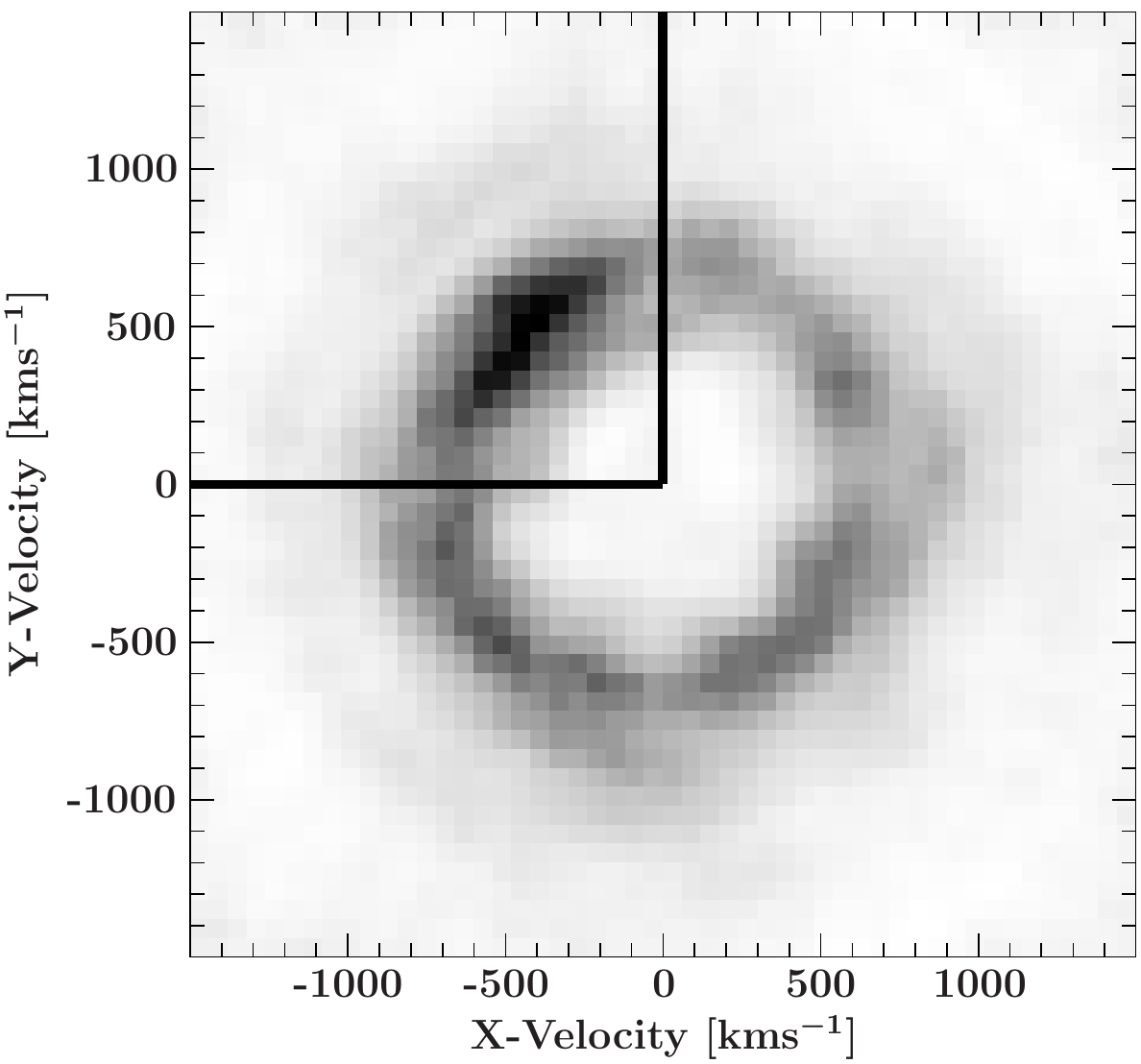}
\caption{Doppler tomogram of the He\,{\sc i}  7065\,\AA line from SDSS\,J0129 with two distinguished areas used for the temperature estimation. The upper left quarter was taken for the bright spot region, whereas the other three quarter were taken for the disc region.}
\label{fig:doppler_area}
\end{center}
\end{figure}
The flux ratios obtained from the line emission strengths in the Doppler tomograms were used to estimate the temperature of the disc and the bright spot. To separate the disc and bright spot region, each Doppler tomogram was divided into two sections. The Doppler tomograms of all systems with only one bright spot were divided into three quarters with only disc emission and one quarter including the bright spot region (see Fig.~\ref{fig:doppler_area}). The quarter with the second bright spot was excluded from the disc region in SDSS\,J1208. For this system only half of the Doppler tomogram was used for the disc region.

Every section was divided into radial bins with a bin size of 60\,km\,s$^{-1}$, starting at the center of the Doppler tomogram ($v_x,v_y$=0,0). The center of the Doppler tomogram is assumed to be the center of the WD primary due to the extreme mass ratios expected in AM\,CVn systems where the orbital velocity of the primary in long period systems is expected to be less than 50\,km\,s$^{-1}$ \citep{mor03, roe06a}. The average flux per radial bin was calculated, which leads to a radial emission profile of the disc and the bright spot. Finally the flux of the radial profile was summed to obtain the full flux in a line. 

To isolate the bright spot emission, the radial emission profile of the disc region was subtracted from the radial emission profile of the bright spot region. The innermost radii ($<$300\,km\,s$^{-1}$) in SDSS\,J1208, SDSS\,J1525 and SDSS\,J0129 were excluded to avoid contamination from the central spike. In SDSS\,J1642 the flux was calculated starting from the center (0\,km\,s$^{-1}$) because no central spike is visible in this system.

\begin{figure*}
\begin{center}
\includegraphics[width=0.99\textwidth]{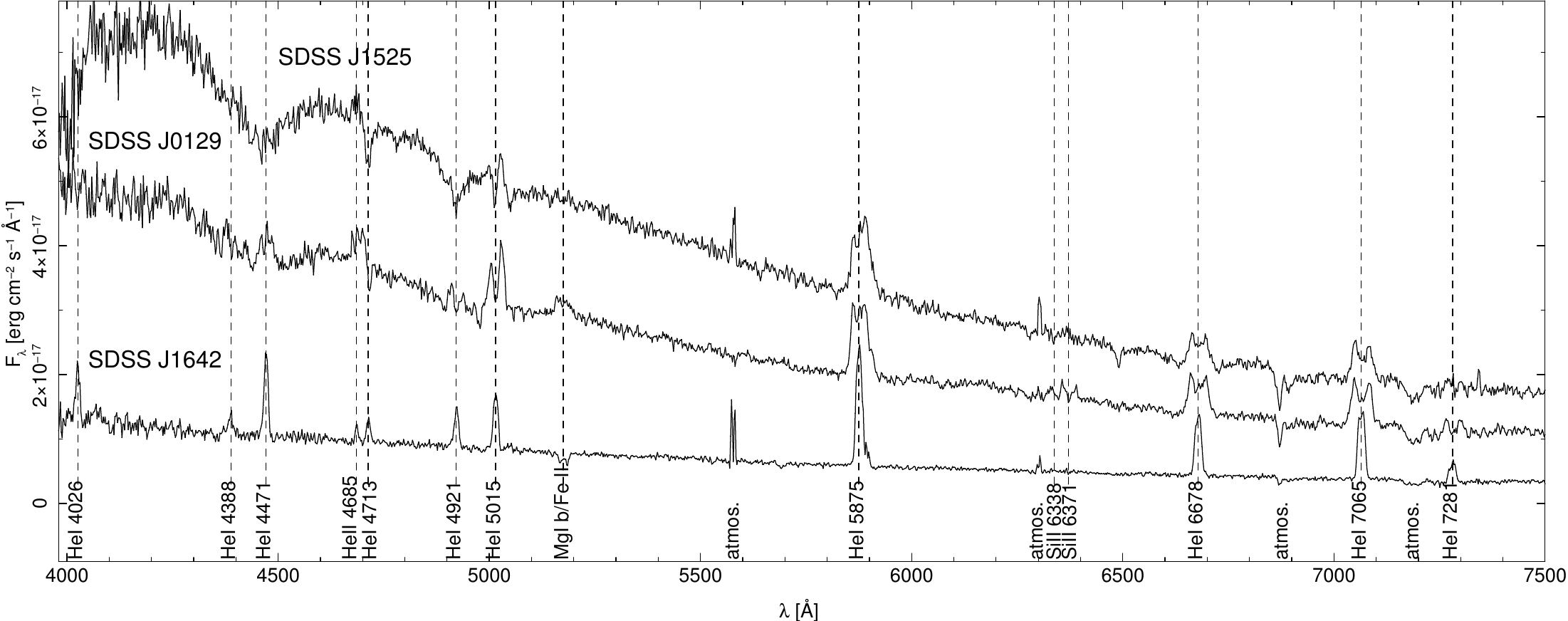}
\caption{Average spectra of SDSS\,J1525 (top), SDSS\,J0129 (middle) and SDSS\,J1642 (bottom) obtained with the GTC. The strongest features are labelled. Each spectrum was flux calibrated but for clarity the flux of SDSS\,J1525 and SDSS\,J1642 was multiplied by $1.5$ and $1.35$ respectively. The flux of SDSS\,J1642 was divided by a factor of $2.5$. Prominent lines are marked with dashed lines.}
\label{gtcaver}
\end{center}
\end{figure*}

The obtained flux ratios were compared to a single slab LTE model with uniform temperature and density. This model is an approximation as temperature and density will change over the disc. \citet{mar91} found overall good agreement between those models and the observations for GP\,Com. They also found a 25\% discrepancy between the models and the observed strengths in the He\,{\sc i} 5015\,\AA\ line. The same discrepancy is found in all four systems (see upper left panel in Fig.~\ref{fig:disc_do} without 25\% correction and upper right panel in Fig.~\ref{fig:disc_do} with 25\% correction). Therefore, in this analysis a 25\% correction is applied to the He\,{\sc i} 5015\,\AA\ line in all systems. Fig.~\ref{fig:disc_do} shows that excluding the He\,{\sc i} 5015\,\AA\ line leads to a degeneracy in the derived temperatures because the remaining flux ratios still allows a dual solution. The measured flux ratios with the He\,{\sc i} 5015\,\AA\ line are needed to break this degeneracy even when including a 25\% uncertainty on the He\,{\sc i} 5015\,\AA\ line itself. Lines further to the blue are not usable as they are either affected by absorption from the accretor or by a very strong central spike. The single slab LTE model is described in full detail in \citet{mar91}. For the temperature estimation a fixed electron density ($n_{\rm e}$=10$^{14}$cm$^{-3}$) and fixed path length ($l$=10$^9$cm) is used, which are values similar to those found in GP\,Com by \citet{mar91}.
\begin{figure}
\begin{center}
\includegraphics[width=0.45\textwidth]{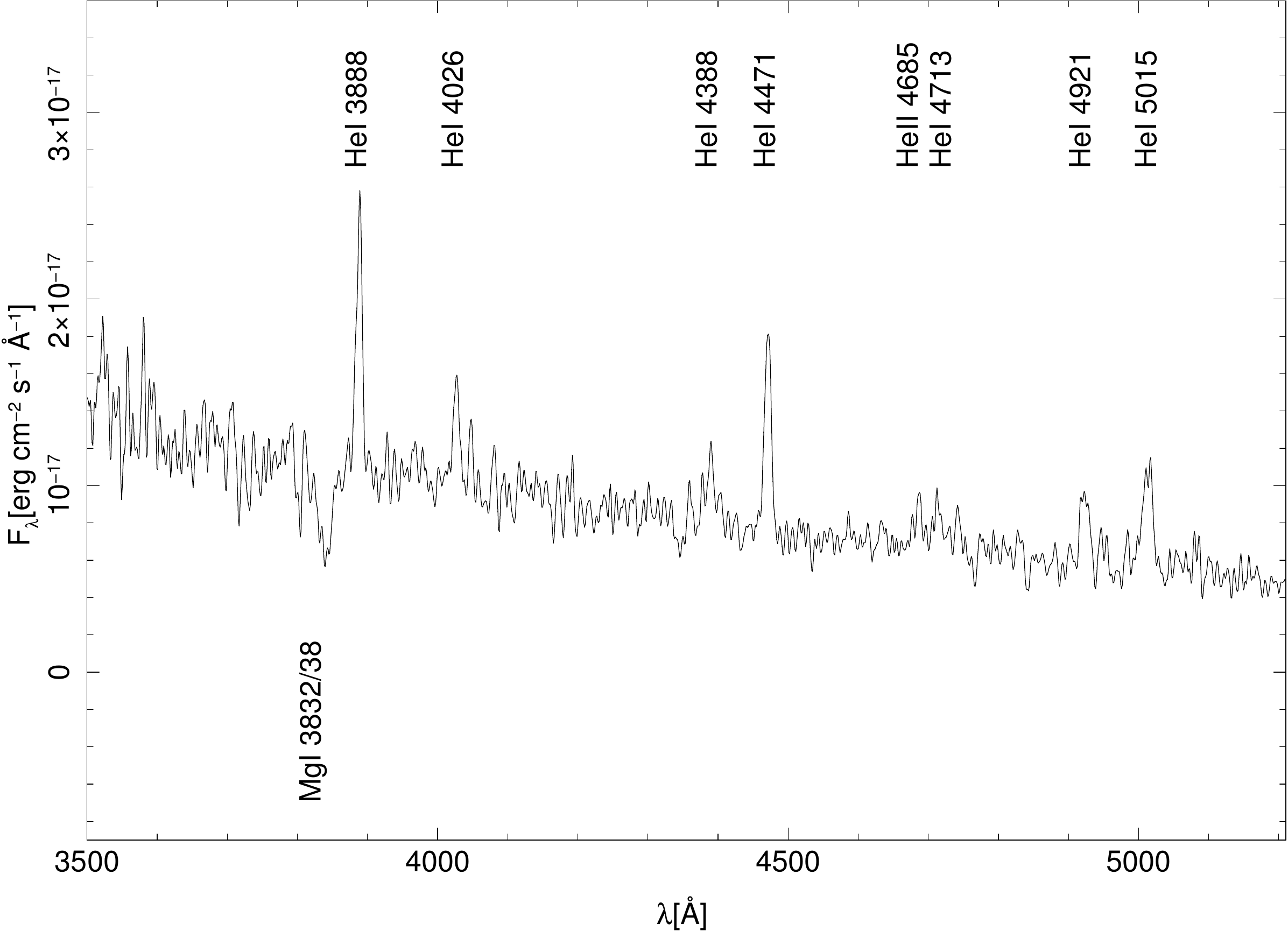}
\caption{Gaussian smoothed spectrum of SDSS\,J1642 obtained with the Hale telescope. Beside the known helium emission lines also the strong Mg\,{\sc i} absorption lines similar to SDSS\,J1208 are present.}
\label{haleaver}
\end{center}
\end{figure}

To estimate the temperature of the disc and bright spot, six different flux ratios of helium lines were used (7065/6678, 7065/5875, 7065/5015, 6678/5875, 6678/5015 and 5875/5015). The output of every Doppler tomogram is set by the targeted $\chi^2$ value. This value has to be set manually and depends on the number of lines used and the quality of the data. A $\chi^2$ value set too high smears out features, whereas a $\chi^2$ value set too low leads to noise artefacts in the reconstructions. The optimal $\chi^2$ is a compromise between goodness of fit and avoiding overfitting the data. This is a common feature of all maximum entropy reconstruction techniques as, such as e.g. also eclipse mapping \citep{hor85}. Setting the targeted $\chi^2$ turned out to be the dominant uncertainty in our method using actual line fluxes to derive temperatures.

To estimate this uncertainty a simple Monte Carlo simulation was used. For SDSS\,J1208 a number of 150 spectra were randomly picked out of the full sample of 150 spectra, allowing for a spectrum to be picked more than once and also allowing for randomly distributed $\chi^2$ values in the Doppler tomograms. The allowed range of the target $\chi^2$ value is taken to be between half the initial value and the estimated best value. Next, 1000 Doppler tomograms were computed, the flux extracted and the standard deviation of all extracted fluxes taken as the error on the fluxes. The same was done for SDSS\,J0129, SDSS\,J1642 and SDSS\,J1525 using 123, 84 and 70 spectra respectively.

The temperature was calculated taking a weighted mean of all intersection points (see Fig.~\ref{fig:disc_do}). The weighting factor for each intersection point was calculated from the region where the calculated flux ratio from the model lies within the error bars of the measured flux ratio. The error on the temperature estimation was calculated as a weighted error, where the error on each flux ratio was taken as the weighting factor. The obtained error is only the statistical error on the temperatures. The change of the assumed column density leads to a systematic error and is described in Sec.~\ref{sec:column}.

\begin{table*}
 \centering
 \caption{Measured equivalent widths in (\AA) and limits of disc emission and photospheric absorption lines.}
  \begin{tabular}{ccccc}
  \hline
   &  \multicolumn{1}{r}{EW (\AA)} &  \multicolumn{1}{r}{EW (\AA)} & \multicolumn{1}{r}{EW (\AA)} & \multicolumn{1}{r}{EW (\AA)} \\
  Line & \multicolumn{1}{r}{SDSS\,J1208} & \multicolumn{1}{r}{SDSS\,J0129} & \multicolumn{1}{r}{SDSS\,J1642} & \multicolumn{1}{r}{SDSS\,J1525} \\
   \hline\hline
  Disc emission     &    &   &    &   \\
  He\,{\sc i}  3888   &  --4.8 $\pm$ 0.2   &  ...$^a$               &   --7.2 $\pm$ 0.3  &   ...$^a$ \\
  He\,{\sc i}  3964   &  --1.9 $\pm$ 0.2   &  ...$^a$               &   --0.3 $\pm$ 0.2  &   ...$^a$ \\
  He\,{\sc i}  4026   &  --4.6 $\pm$ 0.3   &  ...$^a$               &   --4.2 $\pm$ 0.4  &  ...$^a$ \\
  He\,{\sc i} 4388 & --2.3 $\pm$ 0.3 &   \multirow{2}{*}{4.1 $\pm$ 0.7$^{b, c}$} & --1.8 $\pm$ 0.3  &  \multirow{2}{*}{25.4 $\pm$ 0.7$^{b, c}$}  \\
  He\,{\sc i}  4471   &  --7.0 $\pm$ 0.2   &                  &  --13.0 $\pm$ 0.5  &    \\
  He\,{\sc ii} 4685   &  --1.0 $\pm$ 0.2   &  --3.5 $\pm$ 0.3  &   --1.5 $\pm$ 0.4  &   --0.8 $\pm$ 0.2 \\
  He\,{\sc i}  4713   &  --1.2 $\pm$ 0.2   &   0.8  $\pm$ 0.3 &   --3.8 $\pm$ 0.4  &   1.9 $\pm$ 0.3 \\
  He\,{\sc i}  4921   &  --4.0 $\pm$ 0.2   &   2.4  $\pm$ 0.6$^{b}$ &   --7.8 $\pm$ 0.5  &   6.9 $\pm$ 0.5$^{b}$\\
  He\,{\sc i}  5015   &  --7.1 $\pm$ 0.3   &  --7.9 $\pm$ 0.4  &  --13.2 $\pm$ 0.5  &   0.7 $\pm$ 0.3$^{b}$ \\
  He\,{\sc i}  5875   &  --19.8 $\pm$ 0,4  & --24.4 $\pm$ 0.4  &  --56.8 $\pm$ 0.7  &  --16.0 $\pm$ 0.3 \\
  He\,{\sc i}  6678   &  --15.8 $\pm$ 0.5  & --18.8 $\pm$ 0.5  &  --39.4 $\pm$ 0.7  &   --6.7 $\pm$ 0.3 \\
  He\,{\sc i}  7065   &  --24.0 $\pm$ 0.5  & --25.4 $\pm$ 0.5  &  --54.7 $\pm$ 0.7  &  --11.8 $\pm$ 0.4  \\
  He\,{\sc i}  7281   &   --7.9 $\pm$ 0.4  &  --2.8 $\pm$ 0.4  &  --14.7 $\pm$ 0.7  &  $>$--1.3 \\
  Fe\,{\sc ii} 5169       &        X      &  --2.4 $\pm$ 0.4  &  X  &   $>$--0.5  \\
  \smallskip 
  Si\,{\sc ii} 6347/6371  &       X       &  --3.1 $\pm$ 0.4  &  X  &   --0.6 $\pm$0.2 \\
  Absorption lines      &    &   &    &   \\  
  Mg\,{\sc i} 3832/3838   &  3.2 $\pm$ 0.2  &  ...$^a$ &  10.2  $\pm$ 1.7 &  ...$^a$  \\
  Ca\,{\sc ii} 3933$^{d}$ &  $<$0.3 & ...$^a$  & $<$0.7 & ...$^a$ \\
  Si\,{\sc ii} 3853/58/62 &  1.3 $\pm$ 0.2  &  ...$^a$ &  $<$0.7  &   ...$^a$ \\
  Si\,{\sc i} 3905        &  1.6 $\pm$ 0.2  &  ...$^a$ &  $<$0.7  &  ...$^a$  \\
  Mg\,{\sc i}{\sl b}      &  1.2 $\pm$ 0.2  &  ...$^a$ &  4.0 $\pm$ 0.5  &  ...$^a$  \\
     \hline
\end{tabular}
\begin{flushleft}
Lines marked with an X indicate that this line is not detectable in the spectrum obtained\\
$^a$ Spectrum does not extend to this wavelength\\
$^b$ Equivalent width is combination between disc emission and absorption from accretor\\
$^c$ Combined equivalent width of He\,{\sc i} 4388 and He\,{\sc i} 4471\\
$^d$ Ca\,{\sc ii} 3968 was excluded because of an He\,{\sc i} blend
\end{flushleft}
\label{tab:equi}
\end{table*}
\begin{figure}
\begin{center}
\includegraphics[width=0.48\textwidth]{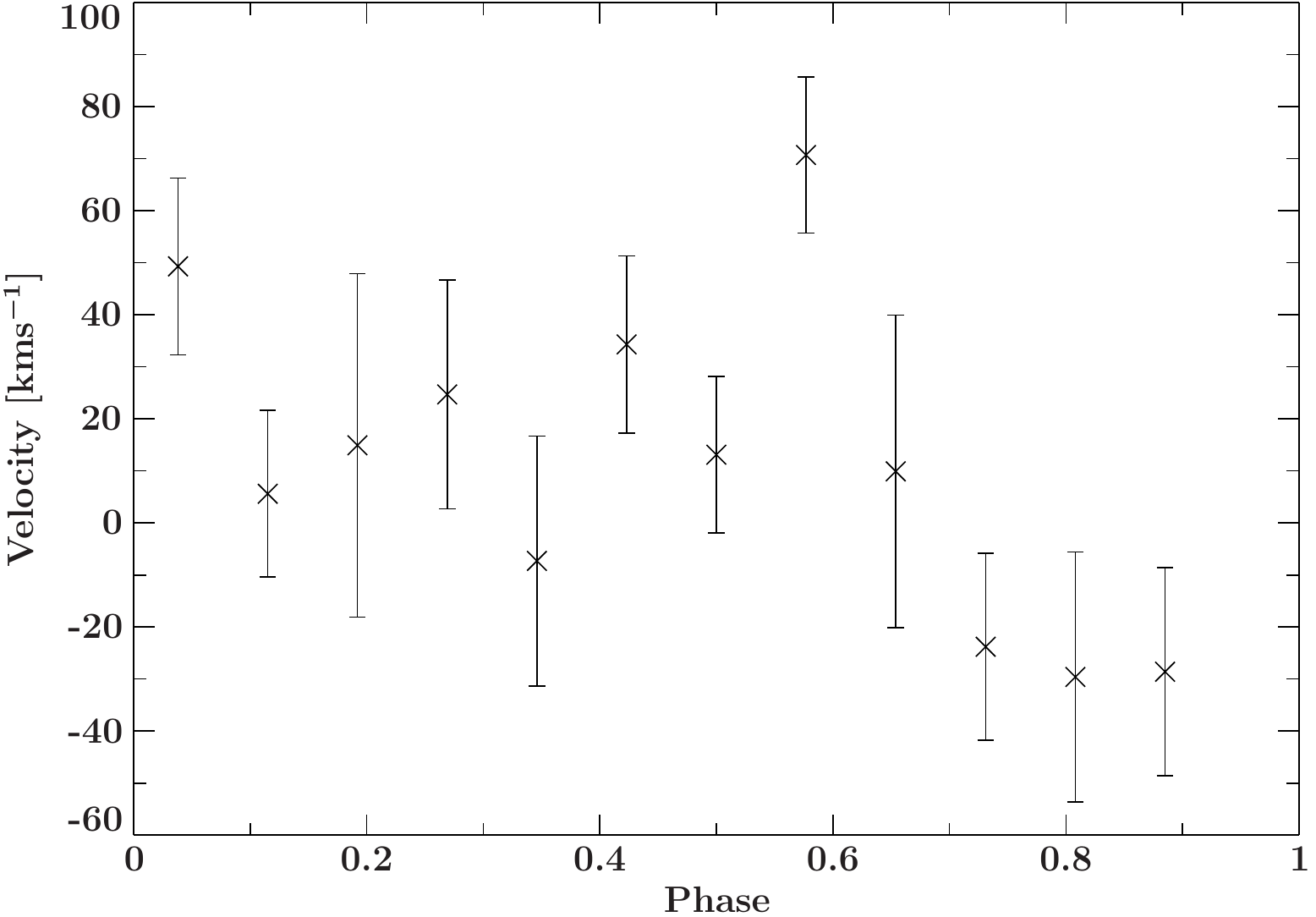}
\caption{Velocity measurement at each phase using a multi-Gaussian fit to the Mg\,{\sc i} lines 3832\,\AA\ and 3838\,\AA\ and the Si\,{\sc i} line 3905\,\AA.} 
\label{fig:velsdss1208}
\end{center}
\end{figure}  

\section{Results}
\subsection{Average spectra}
\subsubsection{Helium lines in absorption and emission}
The average spectrum of SDSS\,J1208 is shown in Fig.~\ref{whtaver}. The strong emission lines of neutral helium and He\,{\sc ii} 4685\,\AA\ are clearly visible. The spectrum looks very similar to the long-period system GP~Com \citep{mar99}. The emission lines show a double-peaked profile originating in the disc, as well as the well-known central-spike feature, which is only observed in AM\,CVn systems \citep{mar99,mor03} and the 65-min orbital period He-rich dwarf nova CSS100603:112253--111037 \citep{bre12}. The He\,{\sc i} 4471\,\AA\ line only shows emission from the central spike. The central-spike feature is known to move with the accreting white dwarf. Therefore, the strong central spike of SDSS\,J1208 was used to measure radial velocities of the accretor and is discussed in Sec.~\ref{sec:radial}.

Fig.~\ref{gtcaver} shows the average spectra of SDSS\,J1525 and SDSS\,J0129. The spectra look very similar. The red part of the spectra is dominated by double peaked emission lines with a small central spike feature in He\,{\sc i} 5875\,\AA\ and no discernible central spike in He\,{\sc i} 6678\,\AA\ and 7065\,\AA. The blue helium lines are dominated by the Stark-broadened absorption lines from the WD accretor. 

The average spectrum of SDSS\,J1642 (Fig.~\ref{gtcaver}) is different from the previous systems, showing only strong emission lines and no sign of an underlying accretor. The double peaks of the individual helium lines are barely resolved and no central spike is seen in this spectrum. 
 \begin{figure*}
   \centering
    \begin{minipage}[b]{0.49\textwidth}
    \includegraphics[width=0.99\textwidth]{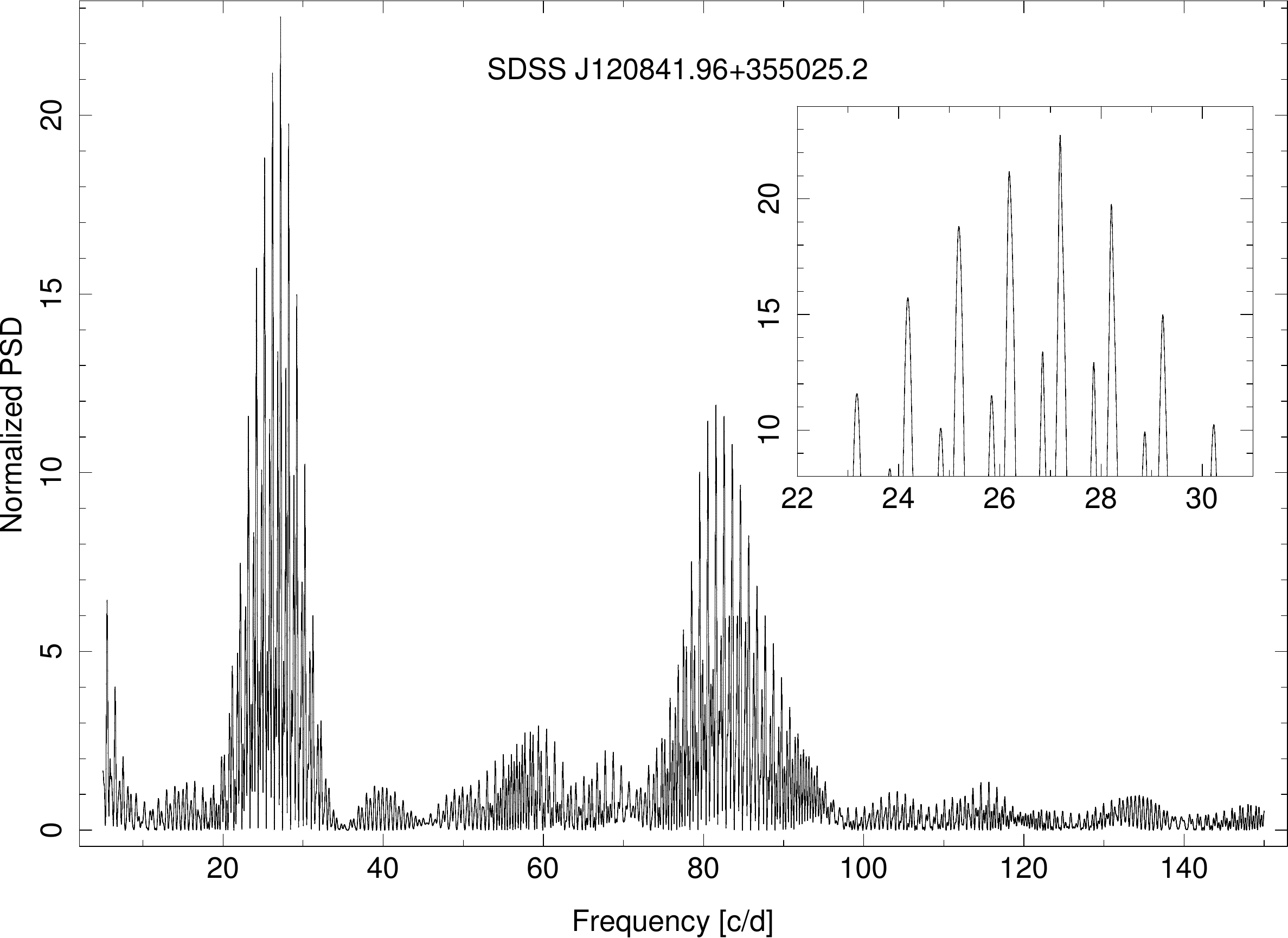}
    \end{minipage}
    \hspace*{0.05cm}
    \begin{minipage}[b]{0.49\textwidth}
    \includegraphics[width=0.99\textwidth]{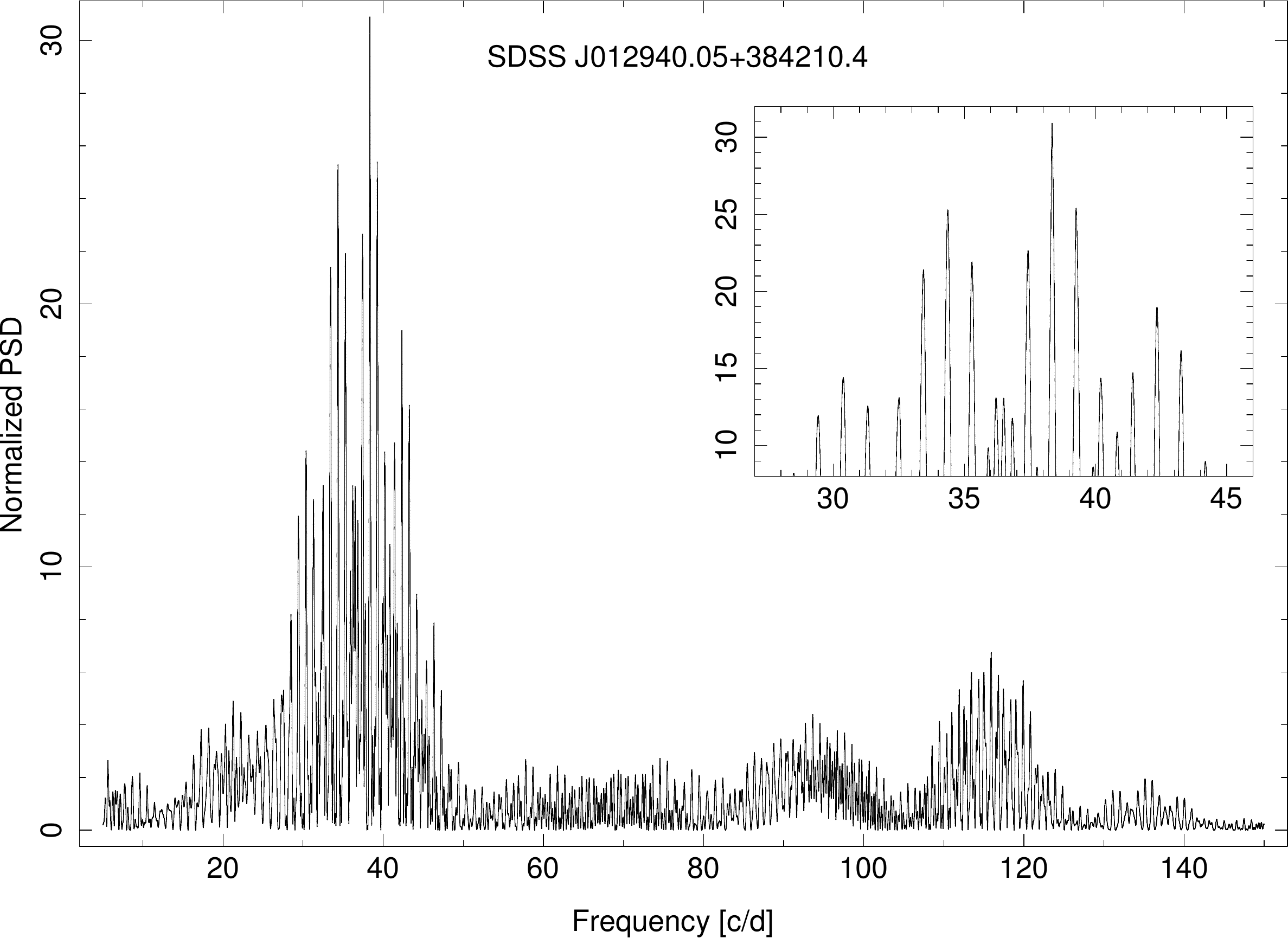}
    \end{minipage}
      \begin{minipage}[b]{0.49\textwidth}
    \includegraphics[width=0.99\textwidth]{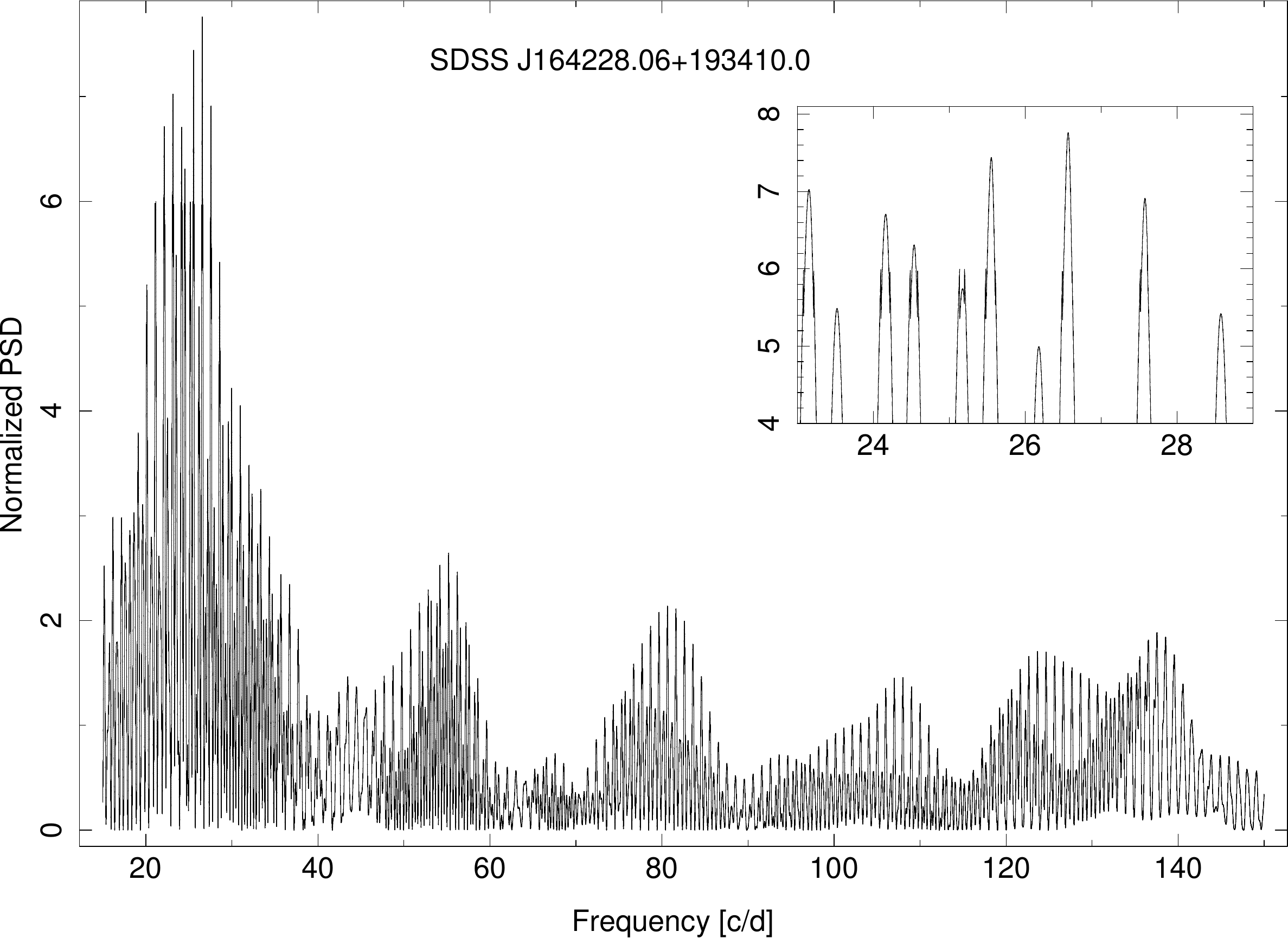}
    \end{minipage}
    \hspace*{0.05cm}
    \begin{minipage}[b]{0.49\textwidth}
    \includegraphics[width=0.99\textwidth]{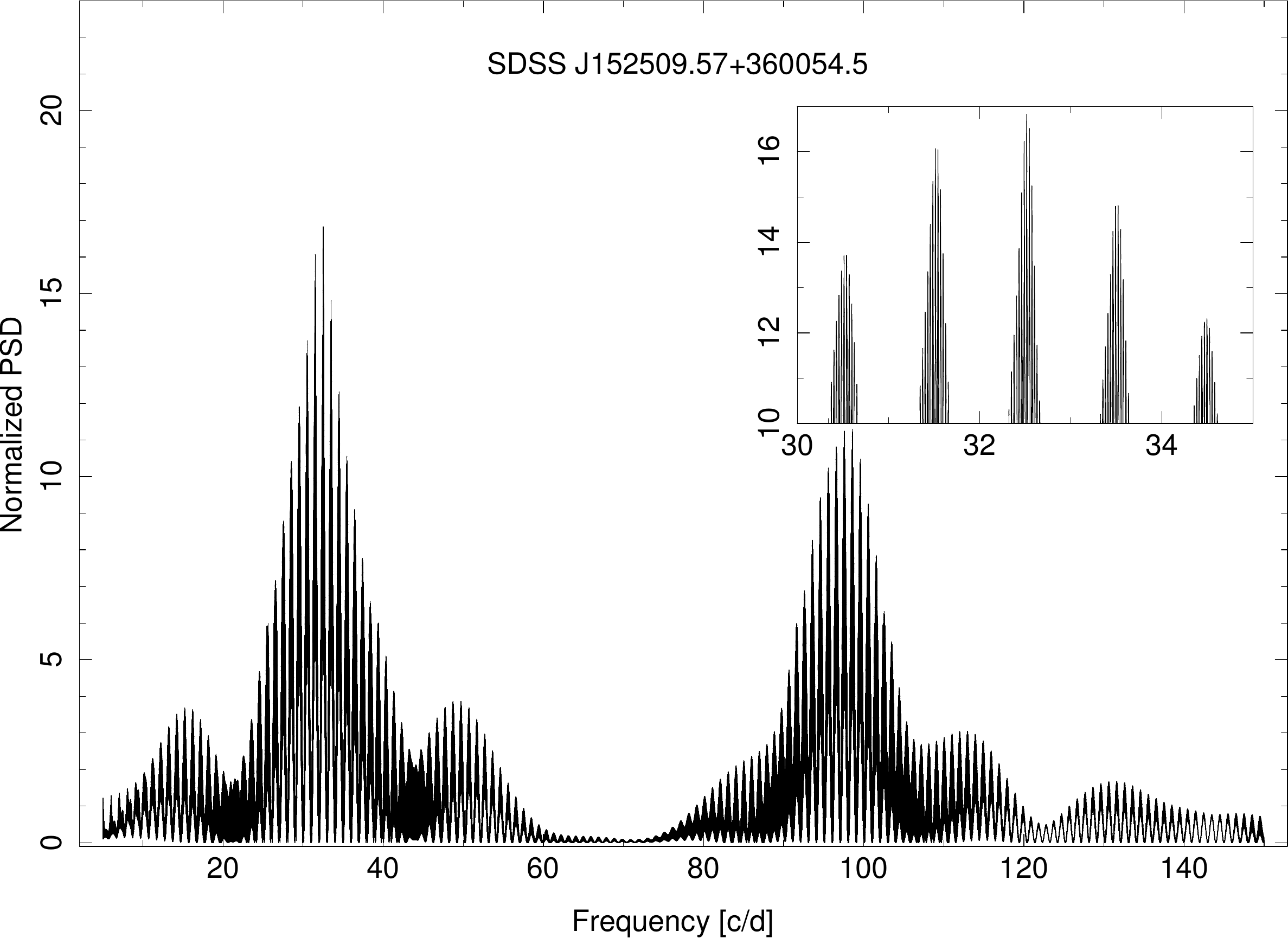}
    \end{minipage}
    \caption{Lomb-Scargle periodograms of the red over violet wing flux ratio. The strongest peak was chosen to be the correct period. Top left: SDSS\,J1208, top right: SDSS\,J0129, lower left: SDSS\,J1642 and lower right: SDSS\,J1525}
    \label{fig:lombscargle}
   \end{figure*}
 \begin{figure}
\begin{center}
\includegraphics[width=0.48\textwidth]{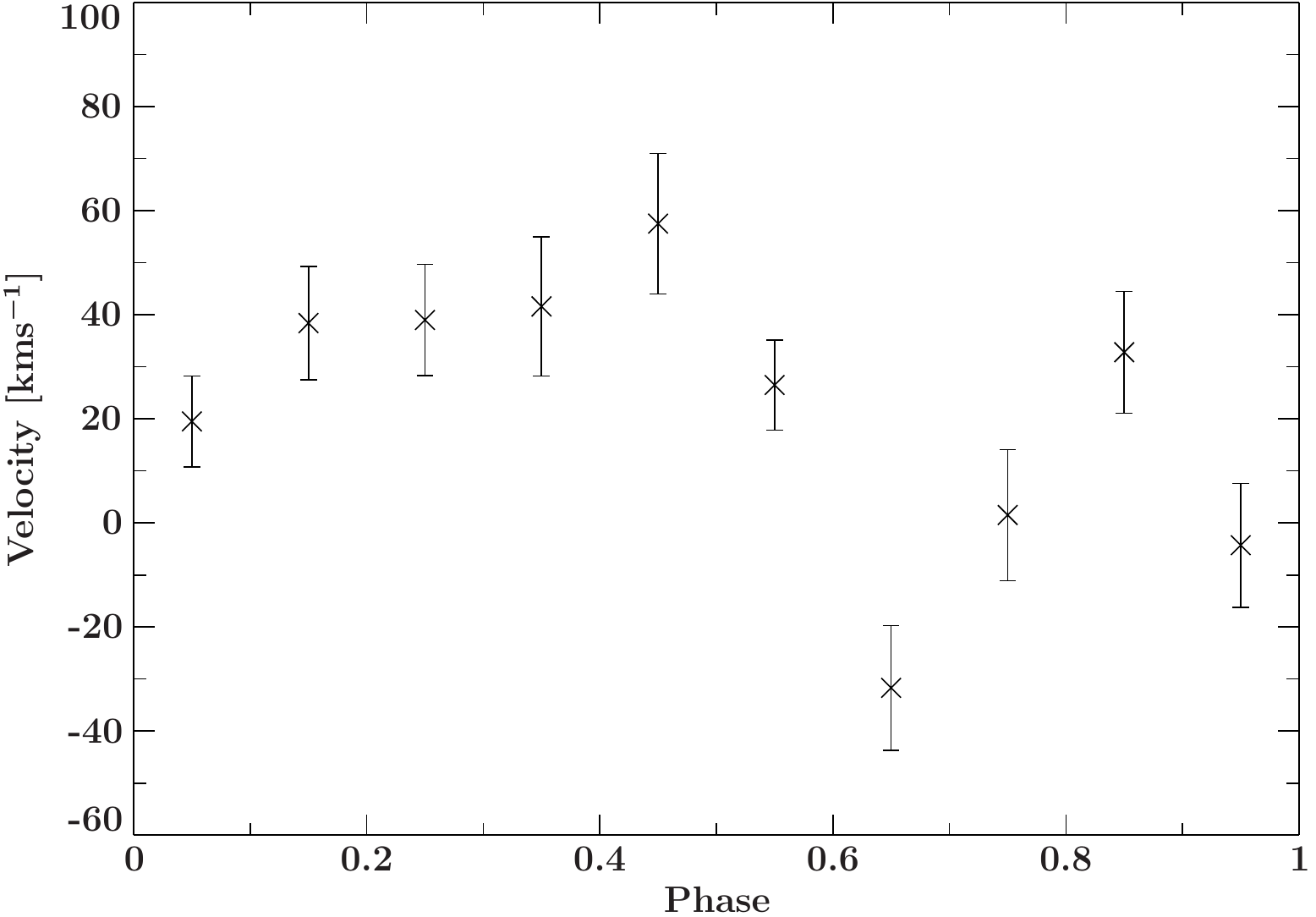}
\caption{Velocity measurement of SDSS\,J1208 at each phase using a multi-Gaussian fit to the helium lines which show only central-spike emission (4026\,\AA, 4471\,\AA, 4921\,\AA, 4685\,\AA\ and 3819\,\AA).} 
\label{fig:velsdss1208_accre}
\end{center}
\end{figure} 
\subsubsection{Emission lines of metals}
Emission lines of various metals including calcium, silicon, iron and nitrogen are observed in quiescent spectra of AM\,CVn systems \citep{rui01,mor03,roe06,roe07a,roe09}. In particular SDSS\,J080449.49+161624.8 (hereafter SDSS\,J0804; Roelofs et al. 2009\nocite{roe09}) shows a very rich emission line spectrum. 

\citet{mar91} predicted Si\,{\sc ii} emission at 6346\,\AA\ and 6371\,\AA\ and Fe\,{\sc ii} emission at 5169\,\AA\ to be the strongest metal lines in helium-dominated optically-thin accretion discs. In principle, the strength of these lines can be used to determine the initial metalicity since their abundance is not supposed to be affected by nuclear synthesis processes during binary evolution. SDSS\,J0129 shows emission of Fe\,{\sc ii} at 5169\,\AA\ and Si\,{\sc ii} at 6347\,\AA\ and 6371\,\AA\ whereas the similar system SDSS\,J1525 shows only weak emission of Si\,{\sc ii} at 6347\,\AA\ and 6371\,\AA. SDSS\,J1208 and SDSS\,J1642 show no metal lines in emission (see Tab.~\ref{tab:equi}).

 \begin{figure*}
\begin{center}
\includegraphics[width=0.95\textwidth]{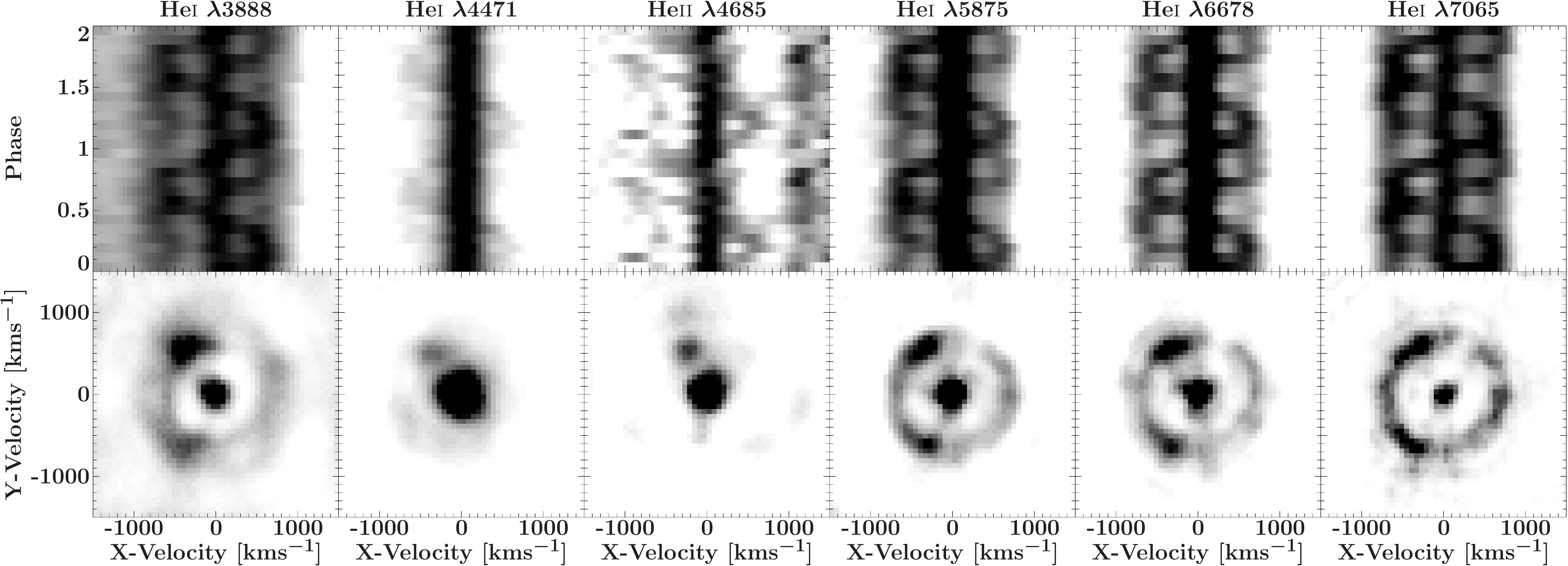}
\caption{Trailed spectra (top row) and maximum-entropy Doppler tomograms (bottom row) of selected He\,{\sc i} and He\,{\sc ii} lines of SDSS\,J1208. Visible is the disc, the central spike as well as both bright spots in some lines. Note that the central spike was saturated to emphasize both bright spots.} 
\label{fig:1208dopplermap}
\includegraphics[width=0.95\textwidth]{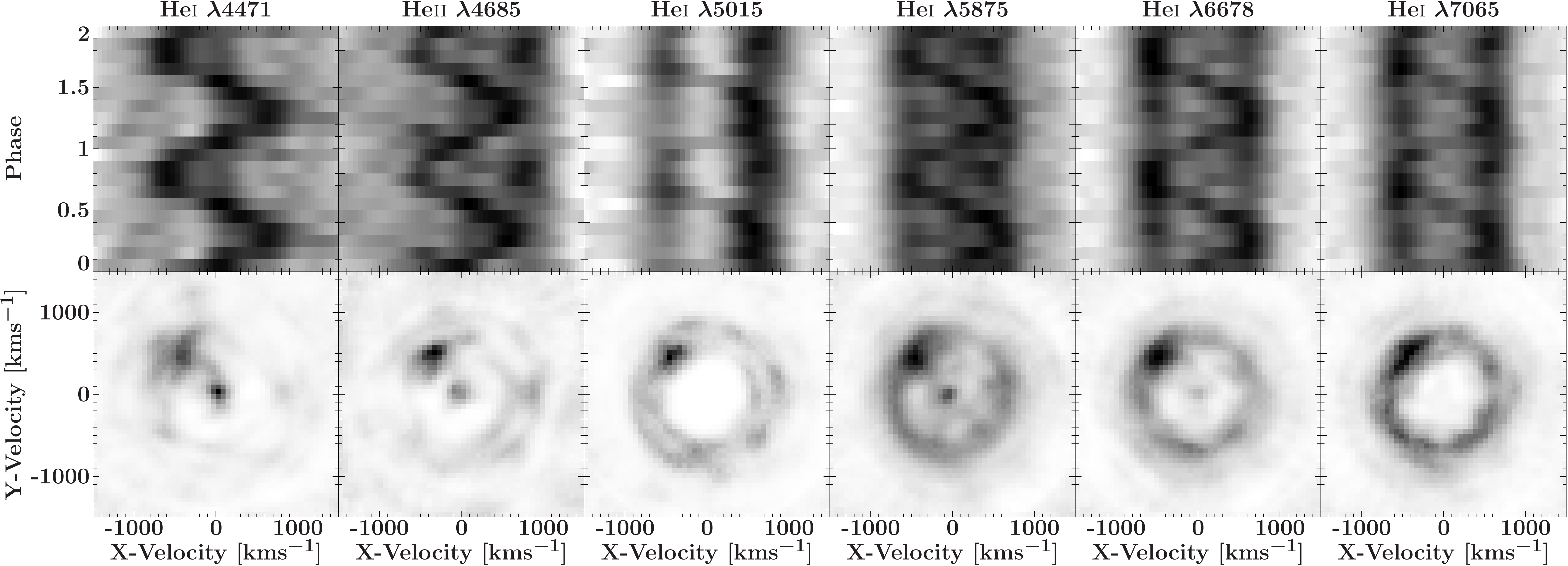}
\caption{Trailed spectra (top row) and maximum-entropy Doppler tomograms (bottom row) of selected He\,{\sc i} and He\,{\sc ii} lines of SDSS\,J0129. Visible is the disc, the bright spot and in some lines the central spike} 
\label{fig:0129dopplermap}
\end{center}
\end{figure*} 

\subsubsection{Absorption lines of metals}
In contrast to metal lines in emission, metal lines in absorption are not common for quiescent spectra of long-period AM\,CVn systems. The only known long-period system in quiescence that shows metal lines in \emph{absorption} is SDSS\,J155252.48+320150.9 (hereafter SDSS\,J1552; Roelofs et al. 2007). \nocite{roe07a}
  \begin{figure*}
\begin{center}
\includegraphics[width=0.7\textwidth]{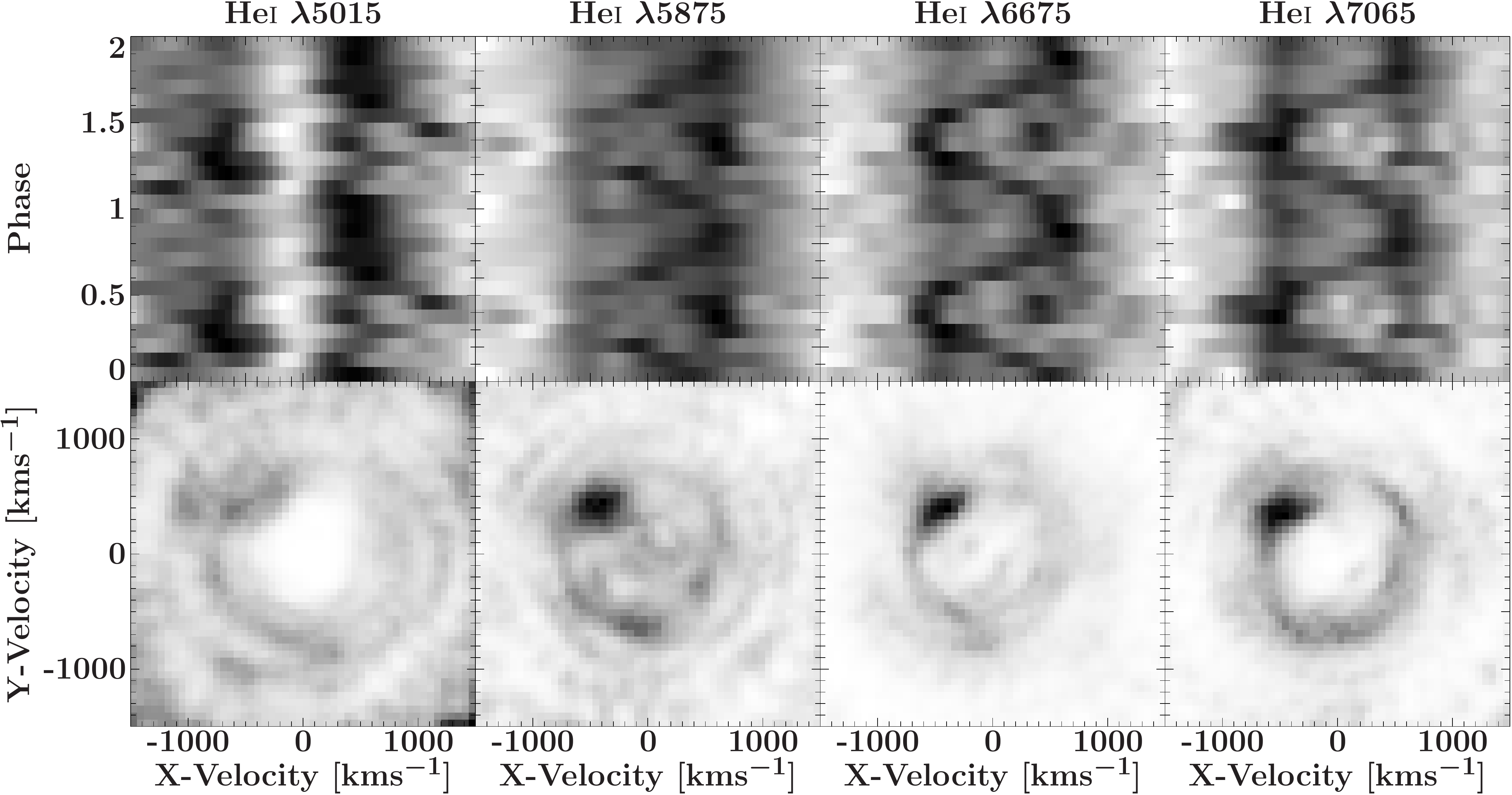}
\caption{Trailed spectra (top row) and maximum-entropy Doppler tomograms (bottom row) of selected He\,{\sc i} lines of SDSS\,J1525. The disc and bright spot are visible.} 
\label{fig:1525dopplermap}
\includegraphics[width=0.7\textwidth]{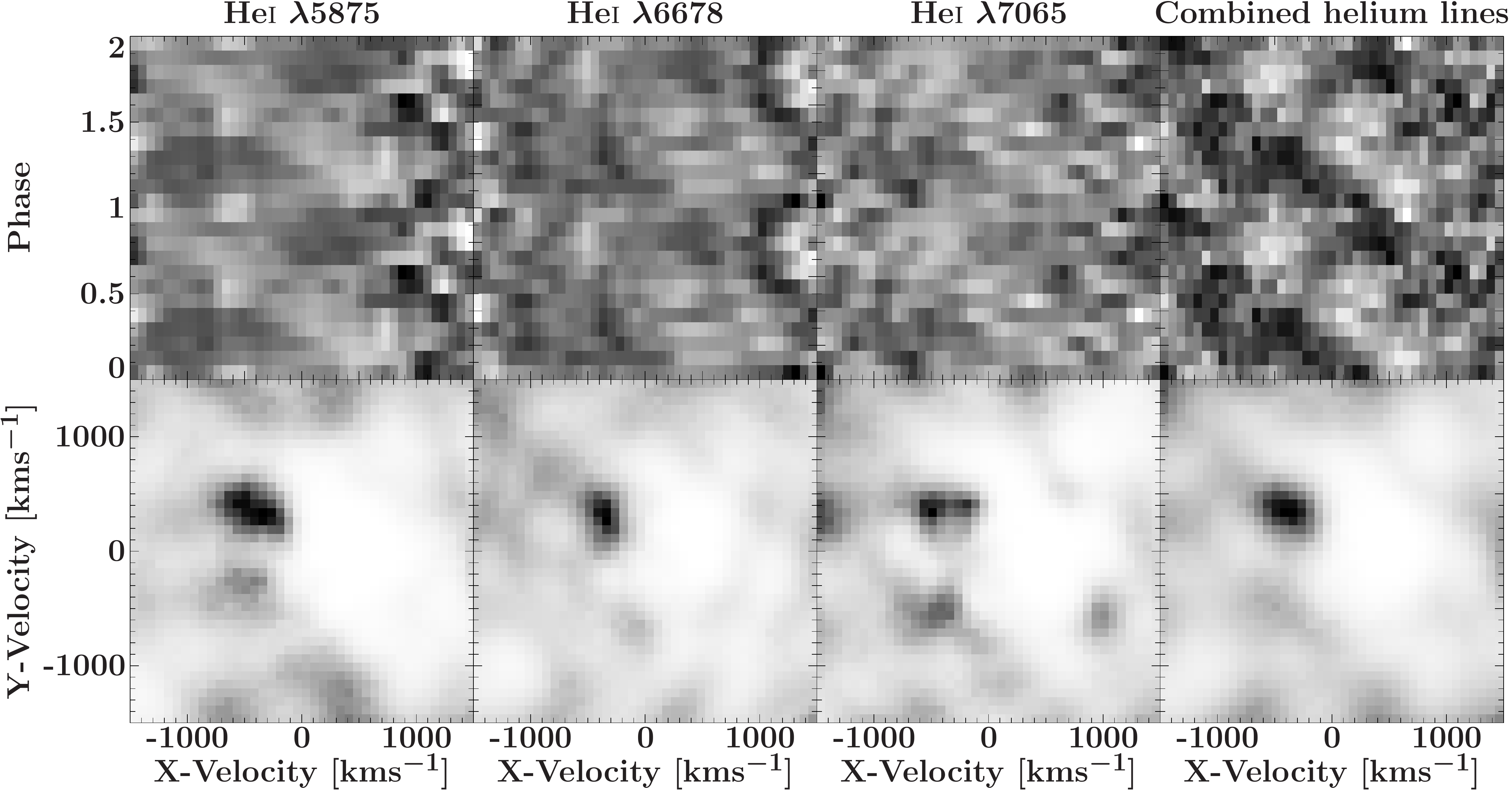}
\caption{Trailed spectra (top row) and maximum-entropy Doppler tomograms (bottom row) of selected and combined helium lines (4471\,\AA, 4713\,\AA, 4921\,\AA, 5015\,\AA, 5875\,\AA, 6678\,\AA, 7065\,\AA, 7281\,\AA) He\,{\sc i} lines of SDSS\,J1642. The average line profile has been divided out to enhance the visibility of the S-wave. Therefore, only the bright spot is visible} 
\label{fig:1642dopplermap}
\end{center}
\end{figure*} 
 
In the systems discussed here, SDSS\,J1208 displays strong absorption lines of Mg\,{\sc i} 3832\,\AA\ and 3838\,\AA\  and Si\,{\sc i/ii} 3853\,\AA, 3856\,\AA, 3862\,\AA\ and  3905\,\AA\,(Fig.~\ref{whtaver}). A single exposure with the Hale telescope revealed the same signature in SDSS\,J1642: strong Mg\,{\sc i} 3832\,\AA\ and 3838\,\AA\ absorption (Fig.~\ref{haleaver}). Both systems also show the Mg\,{\sc i}{\sl b} triplet\,(Fig. 1, 2). \citet{roe07a} found the Mg\,{\sc i} absorption lines as well in the 56-min-period system SDSS\,J1552 but did not detect silicon absorption. 

   \begin{table}
 \centering
 \caption{Overview of measured system parameters of SDSS\,J1208, SDSS\,J0129, SDSS\,J1642 and SDSS\,J1525}
  \begin{tabular}{cccc}
  \hline
  Object & Period (min) &  T$_{\rm disc}$ [K] & T$_{\rm bright\,spot}$ [K] \\
  \hline\hline
  SDSS\,J1208 & 52.96 $\pm$ 0.40 &  10\,800 $\pm$ 400  &  - \\
  SDSS\,J0129 & 37.555 $\pm$ 0.003 & 10\,600 $\pm$ 200 &  27\,800 $\pm$ 900 \\
  SDSS\,J1642 & 54.20 $\pm$ 1.60 &  10\,200 $\pm$ 300  &  - \\
  SDSS\,J1525 & 44.32 $\pm$ 0.18 & -    &  -  \\ 
    \hline
\end{tabular}
\label{tab:period}
\end{table}

To check the origin of the metal absorption lines in SDSS\,J1208, the individual spectra were phase-binned into 13 bins, and multi-Gaussian fits to the Mg\,{\sc i} 3832\,\AA\ and 3838\,\AA\ as well as Si\,{\sc i} 3905\,\AA\ lines were calculated to measure their radial velocity. No periodic shifts are detected within the error bars (Fig.~\ref{fig:velsdss1208}). In long period systems such as SDSS\,J1208 the mass ratio is extreme (M$_2$/M$_1$=$q\le$0.01) and the center of mass is close to the white dwarf. As no periodic velocity shift could be detected (Fig.~\ref{fig:velsdss1208}), the origin of the metal lines can only be the accreting white dwarf or the disc. The latter one is unlikely because the absorption lines are expected to be broadened due to Keplerian velocities in the disc as seen in high state systems such as AM\,CVn itself \citep{roe06a}. Therefore, it is most likely that these lines originate in the accreting WD. 

The presence and strengths of these absorption lines in AM\,CVns were compared to spectra of DBZ WDs. Absorption features of various metal species are known from DBZ WDs but always occur together with the Ca\,{\sc ii} H \& K lines at 3933.66\,\AA\ and 3968.47\,\AA\ (e.g. Dufour et al. 2012)\nocite{duf12}. In order to set a detection limit in equivalent width for calcium, the calcium absorption lines are simulated with a Voigt profile including the resolution and SNR of the spectrum. For a recent work see \citet{duf12}. No sign of calcium in SDSS\,J1208 exceeding an equivalent width of 0.3\,\AA\ and in SDSS\,J1642 exceeding an equivalent width of 0.7\,\AA is detected. A summary of measured equivalent widths and detection limits is given in Tab~\ref{tab:equi}.

\begin{figure*}
   \centering
    \includegraphics[width=0.49\textwidth]{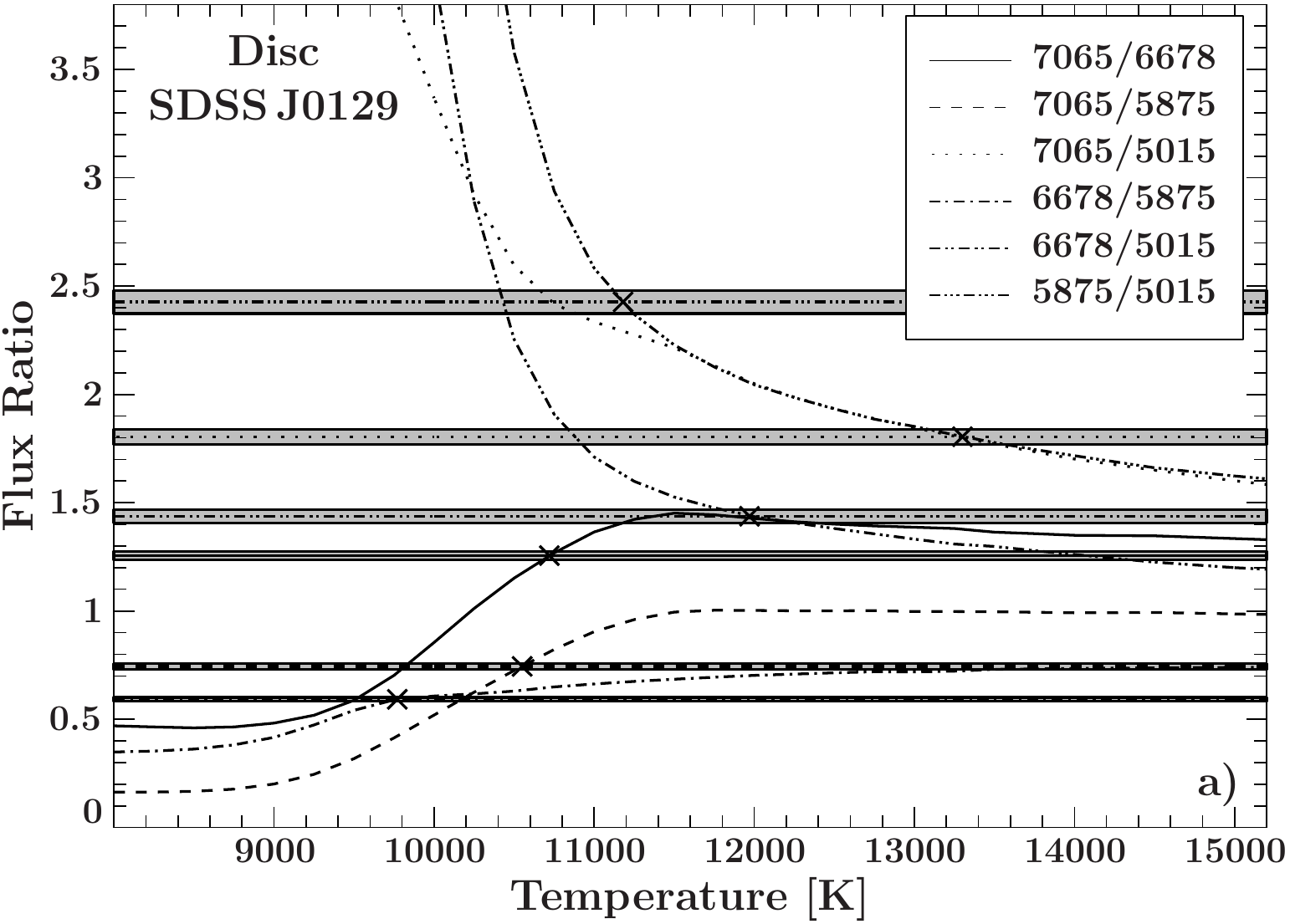}
   \hspace*{0.05cm}
    \includegraphics[width=0.49\textwidth]{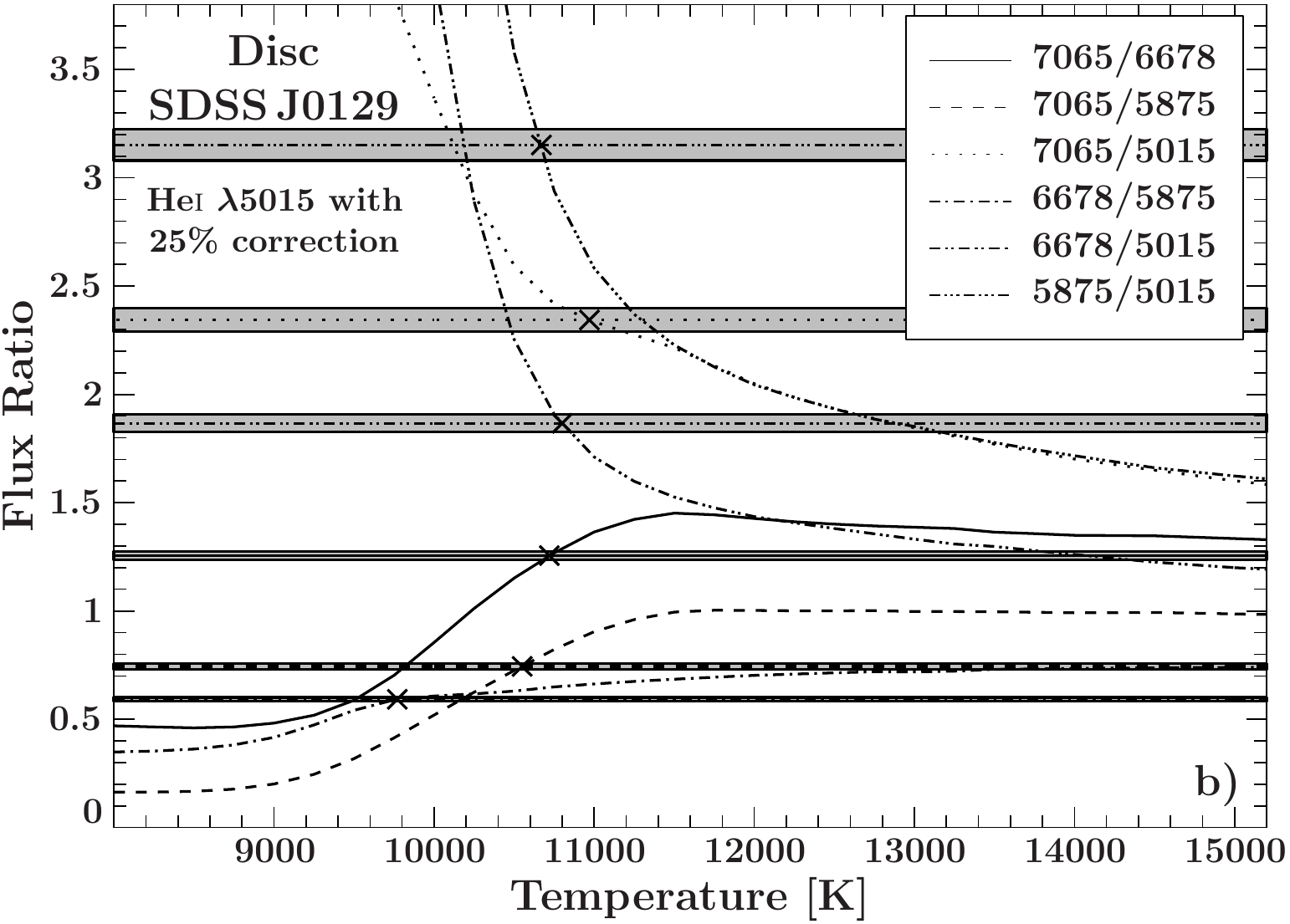}\\
    \includegraphics[width=0.49\textwidth]{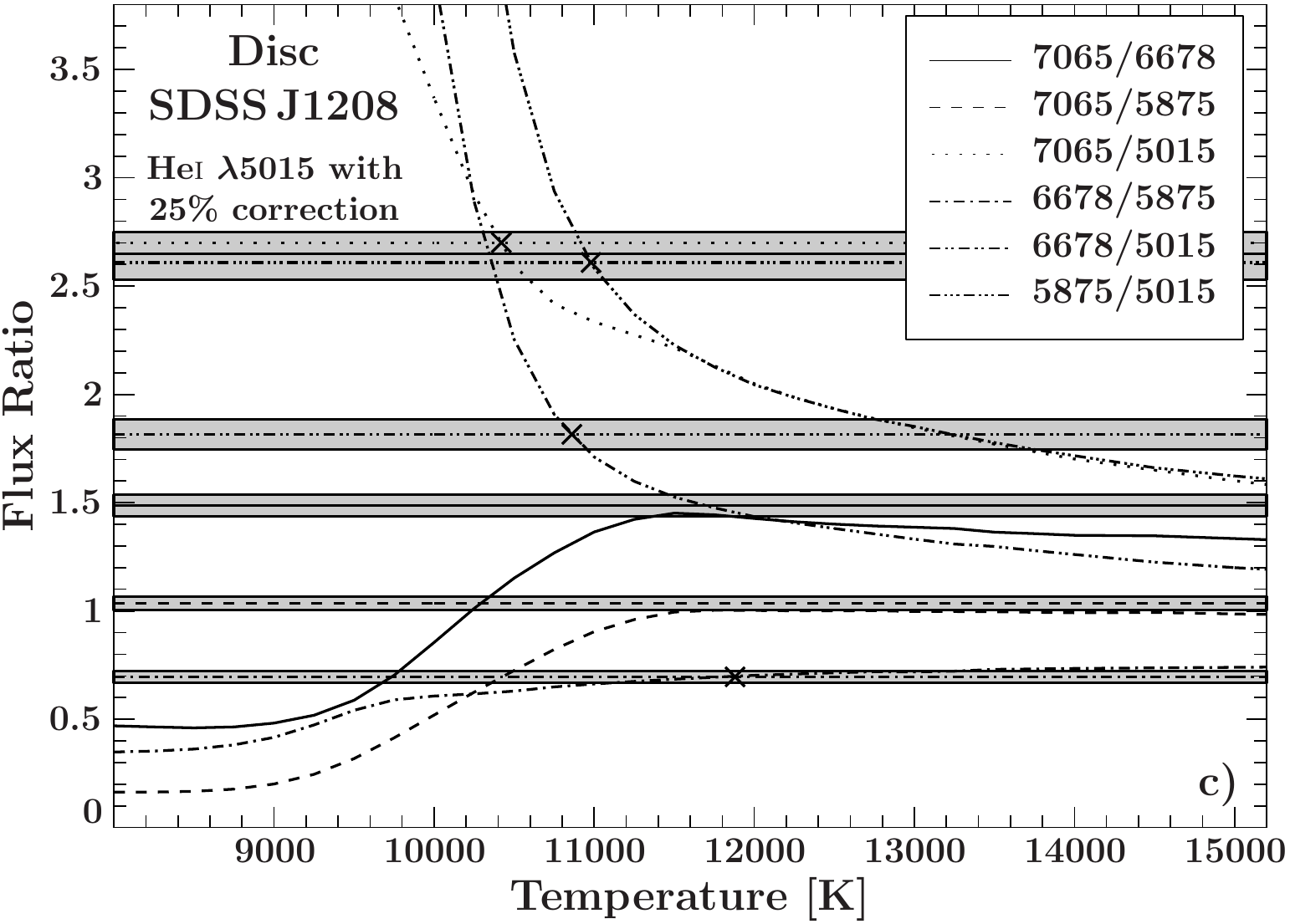}
    \hspace*{0.05cm}
    \includegraphics[width=0.48\textwidth]{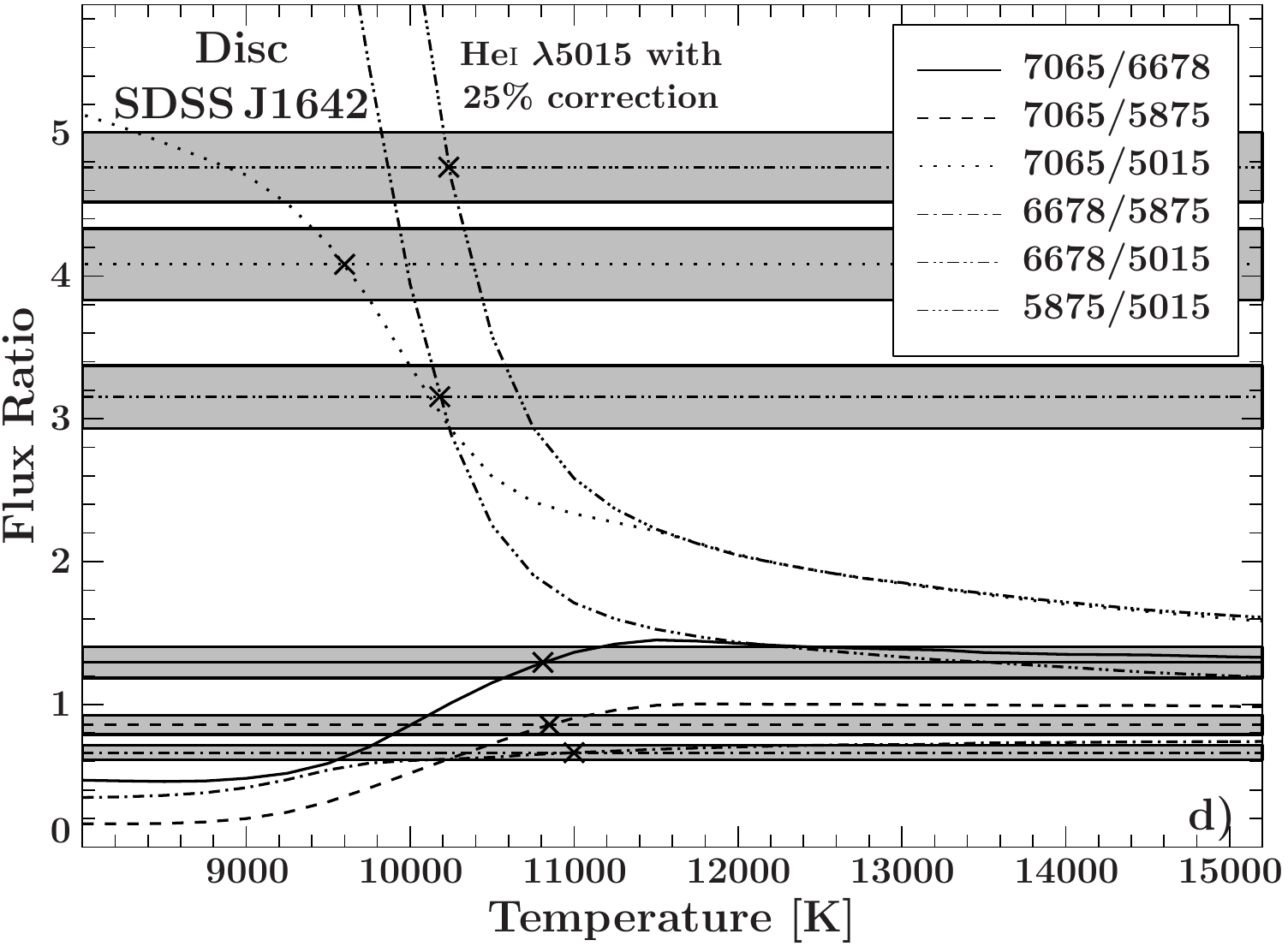}\\
  \begin{minipage}[c]{0.49\textwidth}
   \includegraphics[width=0.99\textwidth]{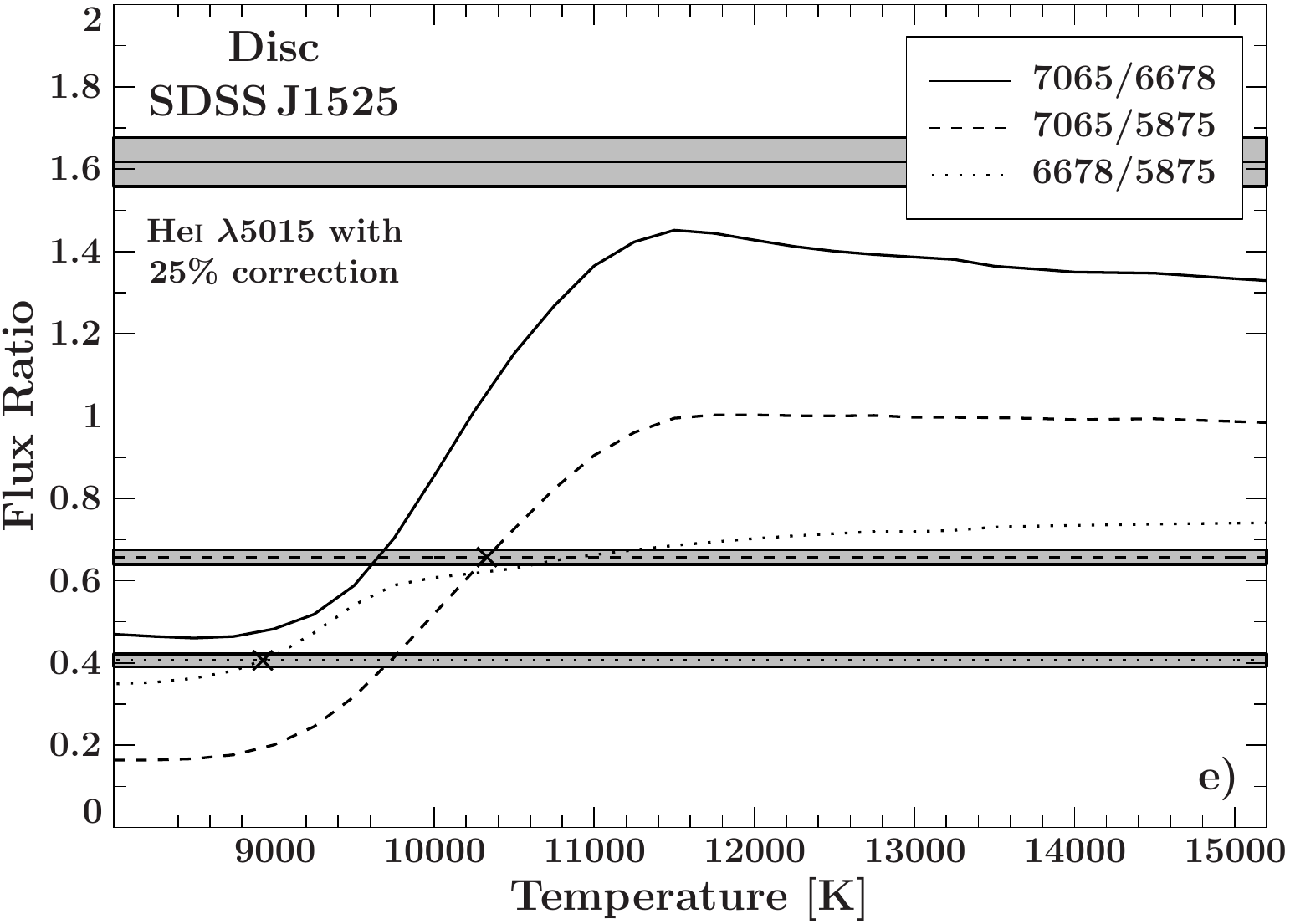}
    \end{minipage}
   \hspace*{0.05cm}
  \begin{minipage}[c]{0.49\textwidth}
         \caption{Comparison between observed disc emission (horizontal lines) and calculated flux ratios for a LTE slab model with a fixed path length of $l$=10$^9$cm and density of 10$^{14}$cm$^{-1}$ for different helium lines of the bright spot emission region (curved lines). The horizontal lines correspond to the measured flux ratios with the error on the flux ratios shaded in grey around the horizontal line. The intersection points (x) for each line are indicated. \citet{mar91} found a disagreement of 25$\%$ between observation and model for the He\,{\sc i} 5015\,\AA\, line. In the upper right hand panel this correction was applied to the flux ratios for SDSS\,J0129. In the upper left hand panel the original flux ratios for SDSS\,J0129 are shown. A better agreement compared to other lines including the correction was found. The other three panels show the flux ratios, models with intersection for the other three systems (c) SDSS\,J1208 with corrected He\,{\sc i} 5015\,\AA\ line, d) SDSS\,J1642 with corrected He\,{\sc i} 5015\,\AA\ line, e) SDSS\,J1525 with corrected He\,{\sc i} 5015\,\AA\ line).}
\label{fig:disc_do}
   \end{minipage} 
\end{figure*}
\subsection{Spectroscopic orbital periods}\label{sec:perioddet}
Using the method described in Sec.~\ref{sec:period_det} orbital periods of 52.96$\pm$0.40\,min for SDSS\,J1208, 37.555$\pm$0.003\,min for SDSS\,J0129, 54.20$\pm$1.60\,min for SDSS\,J1642 and 44.32$\pm$0.18\,min for SDSS\,J1525 are found (see Tab.~\ref{tab:period} and Fig.~\ref{fig:lombscargle}). The spectra for every system were folded onto the obtained period and phase-folded spectra and Doppler tomograms were computed. These are discussed in Section~\ref{sec:doppler}.

All systems, with the exception of SDSS\,J1642, show significant peaks in the periodogram with a high power spectral density (Fig.~\ref{fig:lombscargle}). SDSS\,J1642 shows a weaker bright spot leading to a less significant power spectral density. The highest peak in the periodogram was identified as the orbital period and is given in Table~\ref{tab:period}. A second peak with a slightly longer period shows a similar absolute power spectral density. This peak corresponds to a period of 56.35\,min. 

\subsection{Radial velocities of the accretor in SDSS\,J1208}\label{sec:radial}
The central-spike feature is known to move with the accreting white dwarf and can therefore be used to measure the radial velocity curve and the real phase of the accreting white dwarf. This information in combination with the phase of the bright spot helps to establish the mass ratio of the system \citep{roe06a, mor03, mar99}. 
  
SDSS\,J1525, SDSS\,J0129 and SDSS\,J1642 show only weak or no central spikes and cannot be used to measure radial velocities of the accretor. SDSS\,J1208 on the other hand shows a very strong central spike with some lines strongly dominated by the central spike (e.g. He\,{\sc i} 4471\,\AA, Fig.~\ref{fig:1208dopplermap}). For this line, the individual spectra were phase-binned into 10 bins and a Gaussian fit was calculated to measure the radial velocity. No periodic variations of the central spike within the error bars (Fig.~\ref{fig:velsdss1208_accre}) is detected, which is expected as the velocity amplitude is supposed to be $K\leq$10\,km\,s$^{-1}$.

\begin{figure*}
    \includegraphics[width=0.49\textwidth]{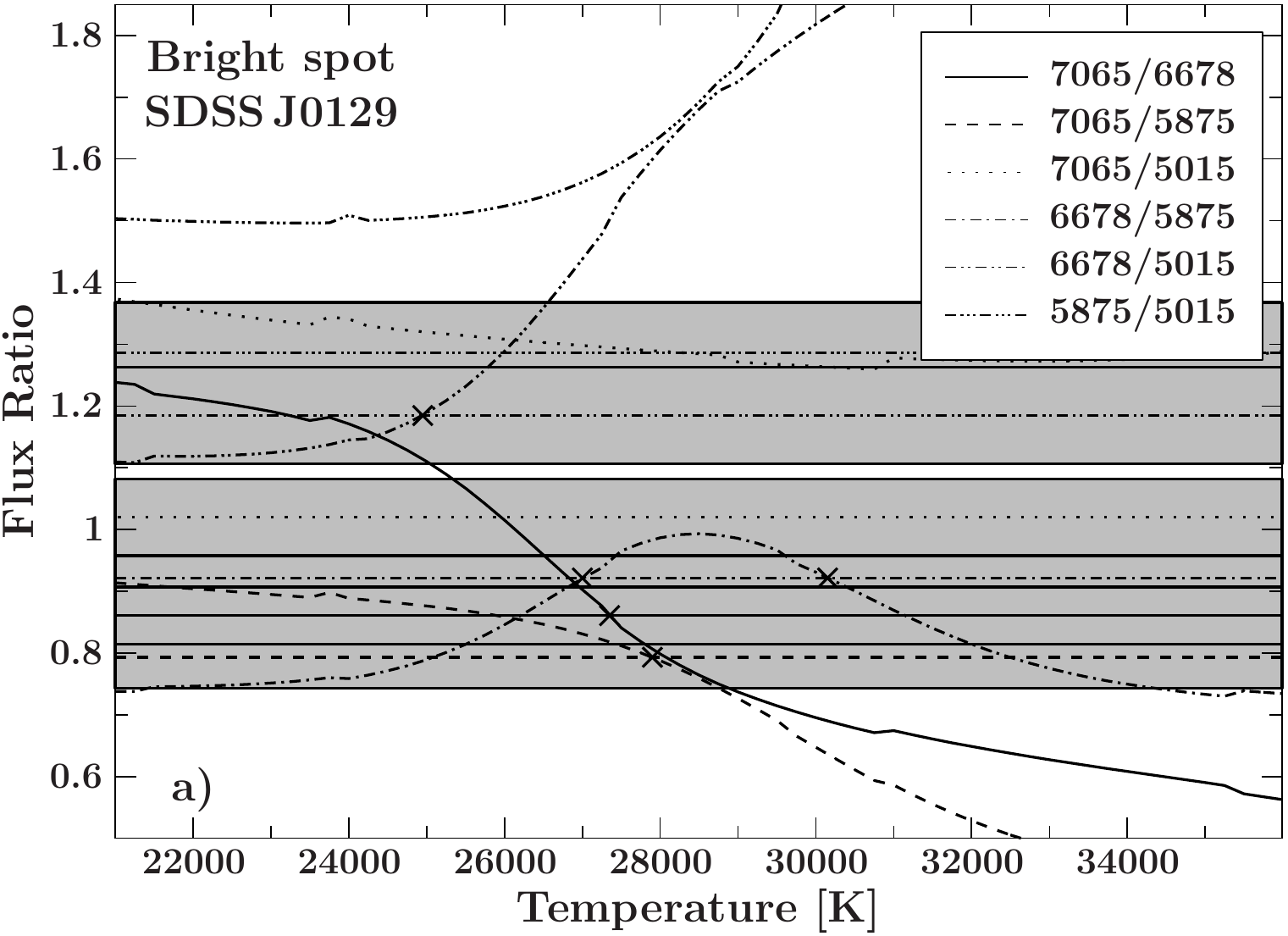}
   \hspace*{0.05cm}
    \includegraphics[width=0.49\textwidth]{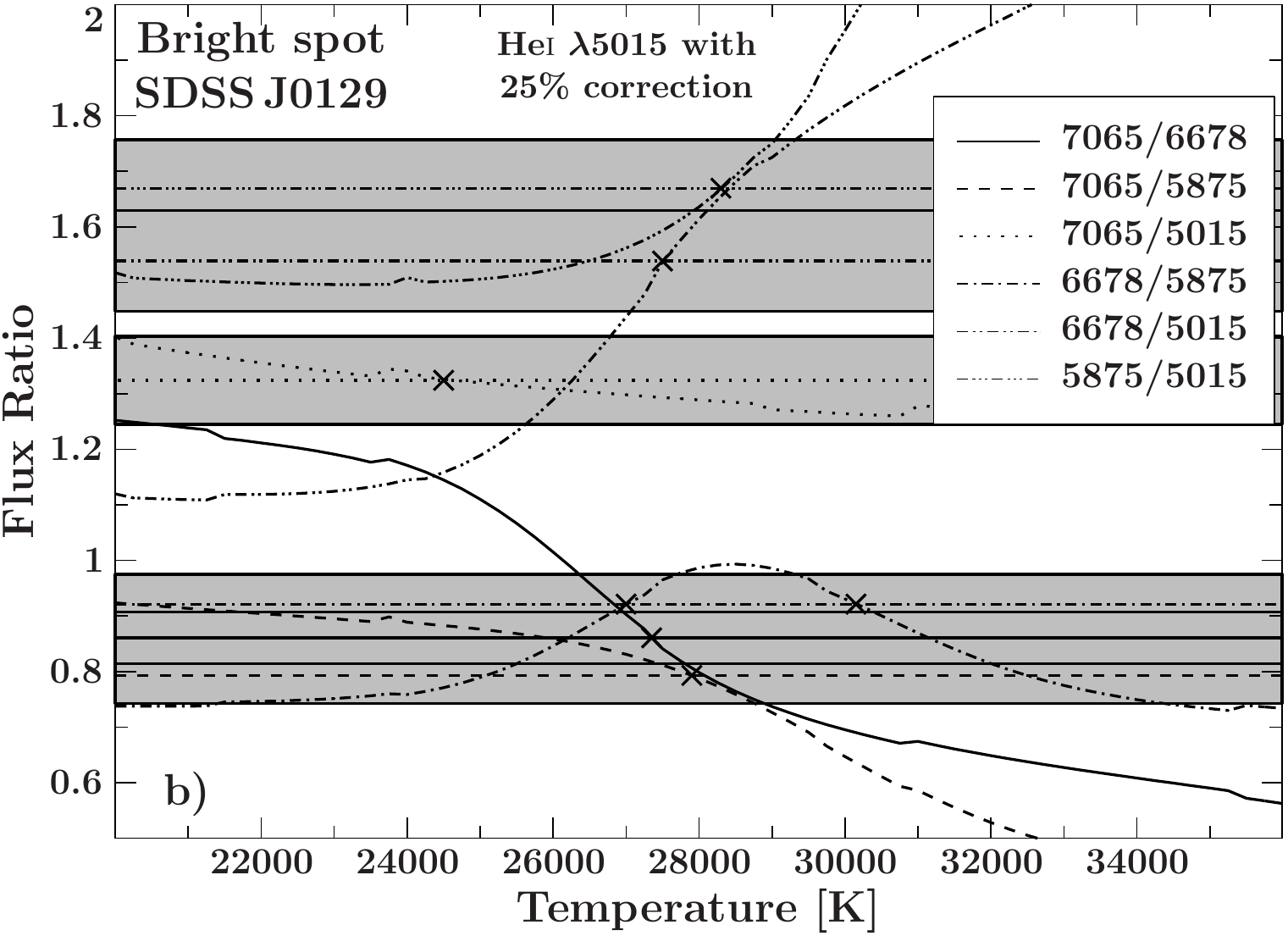}
    \includegraphics[width=0.49\textwidth]{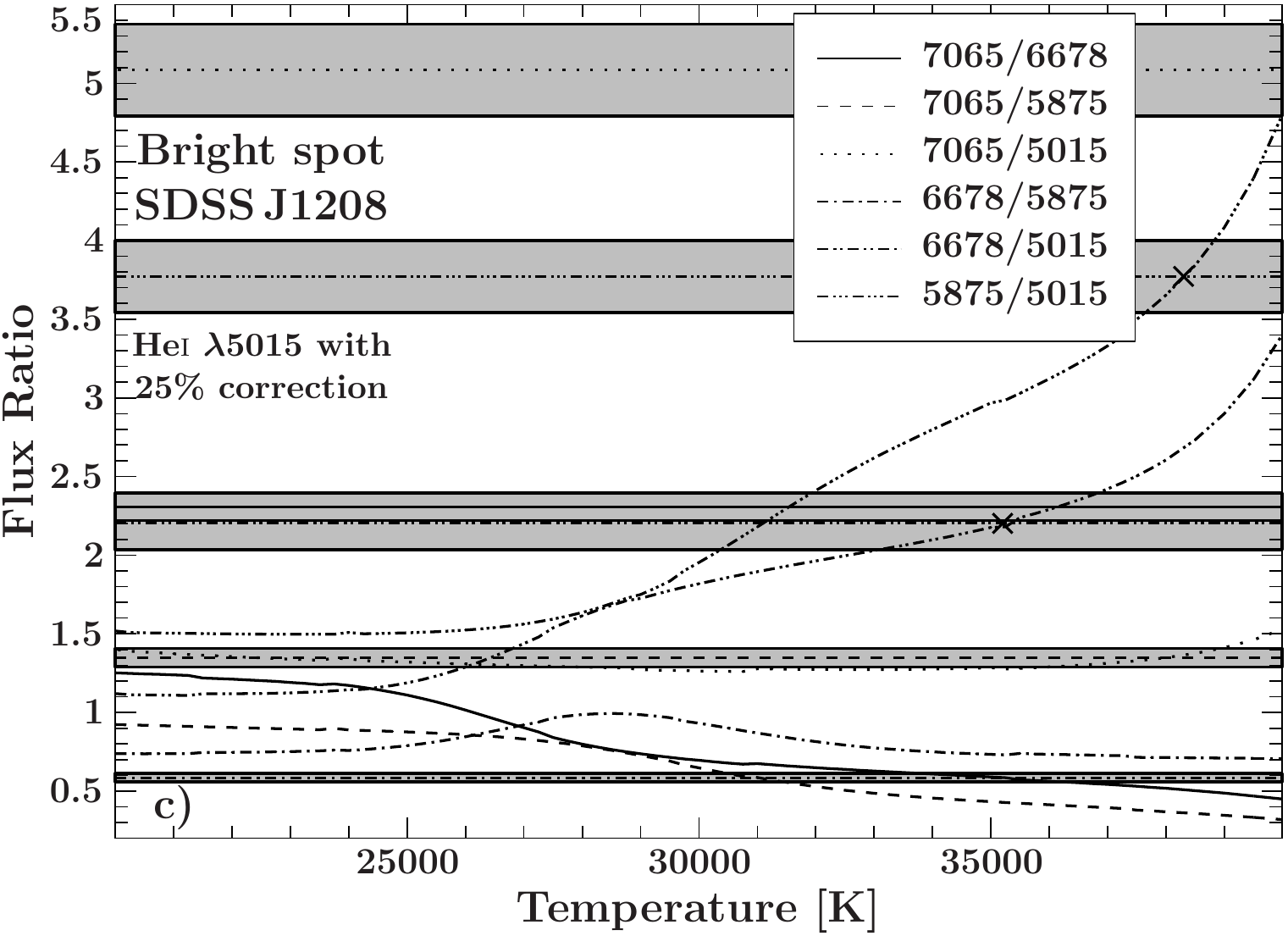}
    \hspace*{0.05cm}
    \includegraphics[width=0.49\textwidth]{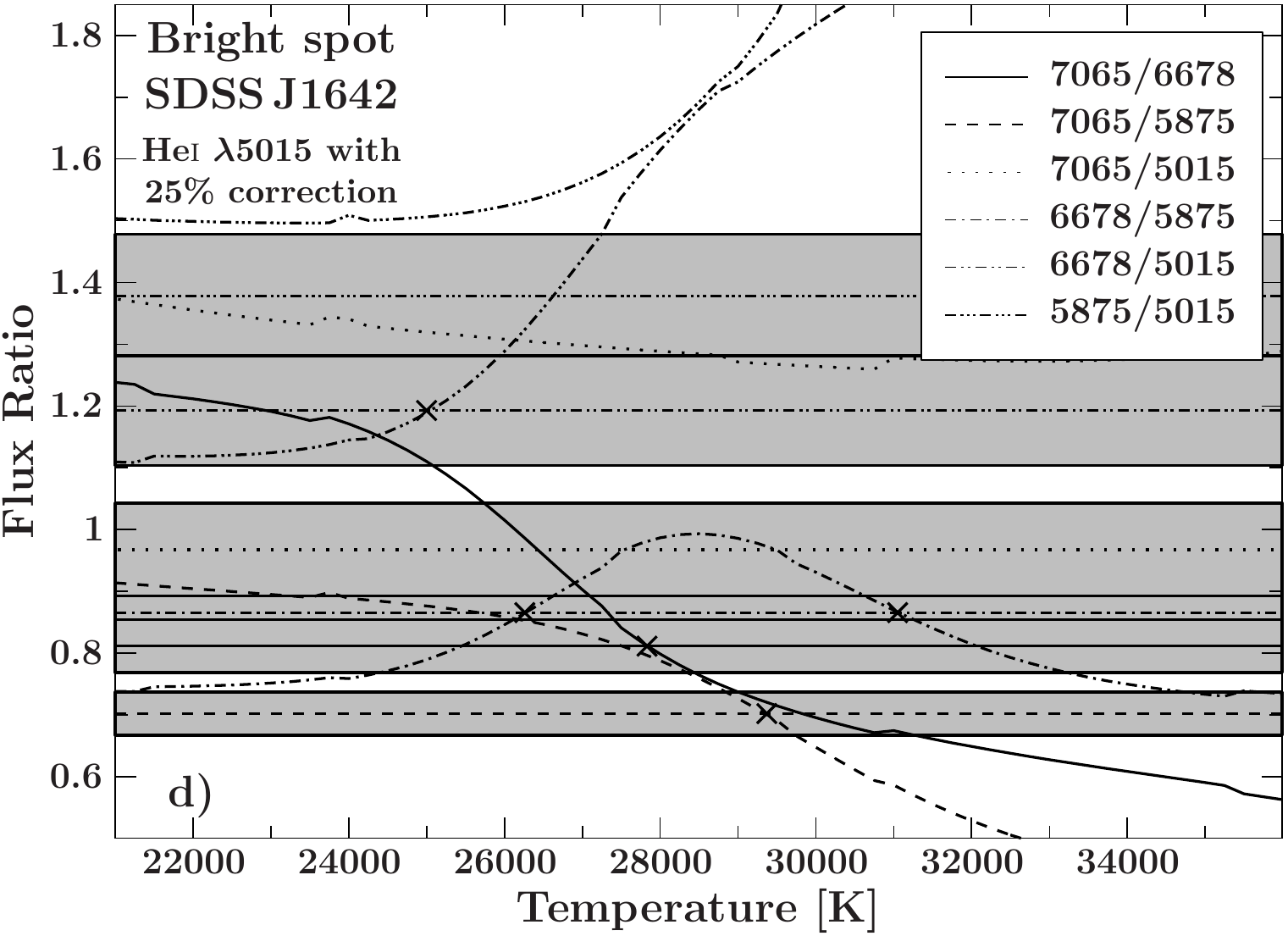}
    \begin{minipage}[c]{0.49\textwidth}
\includegraphics[width=0.99\textwidth]{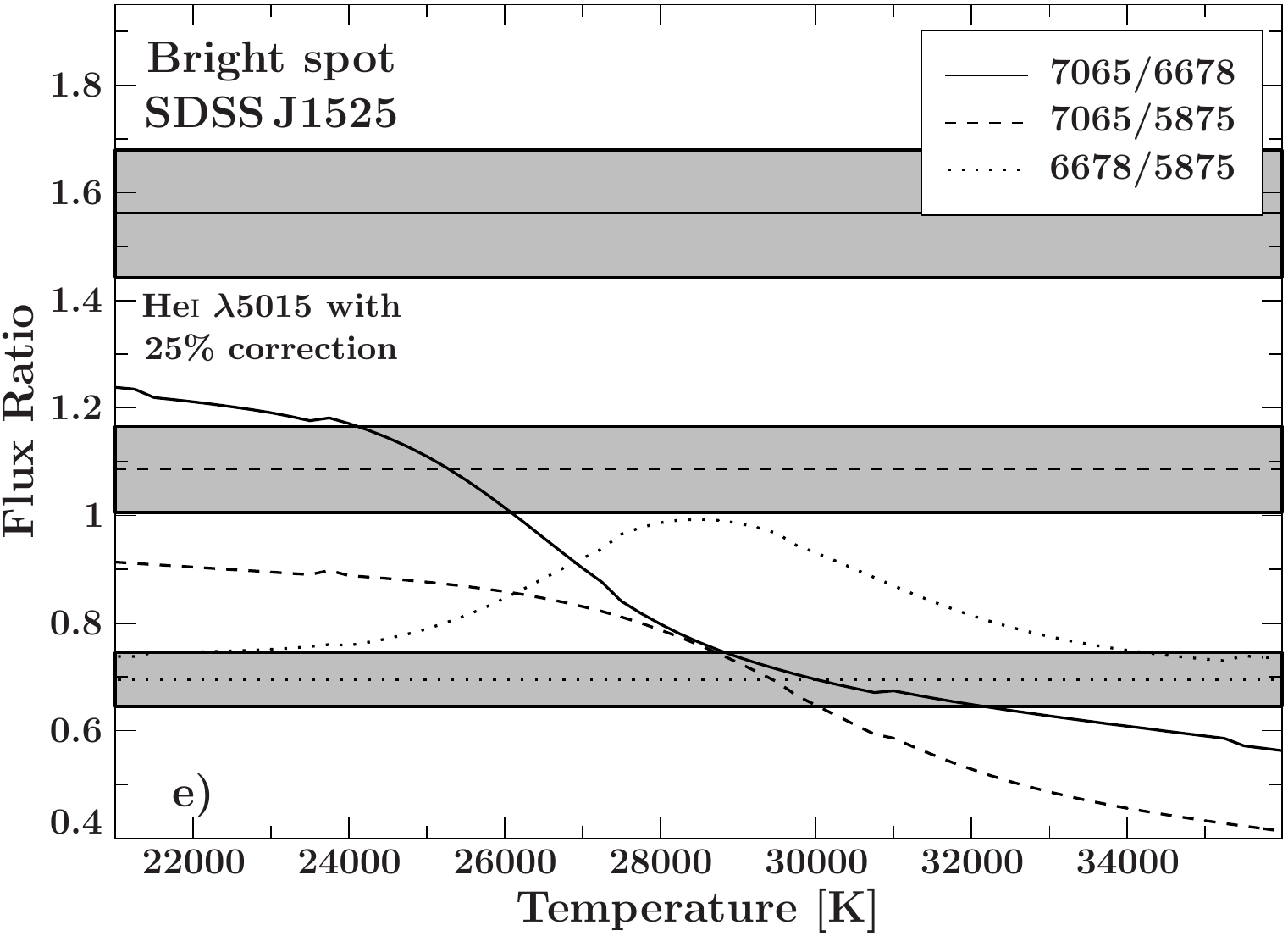}
 \end{minipage}
  \hspace*{0.05cm}
  \begin{minipage}[c]{0.49\textwidth}
    \caption{Comparison between observed bright spot emission (horizontal lines) and calculated flux ratios for a LTE slab model with a fixed path length of $l$=10$^9$cm and density of 10$^{14}$cm$^{-1}$ for different helium lines of the bright spot emission region (curved lines). The horizontal lines correspond to the measured flux ratios with the error on the flux ratios shaded in grey around the horizontal line. The intersection points (x) for each line are indicated. \citet{mar91} found a disagreement of 25$\%$ between observation and model for the He\,{\sc i} 5015\,\AA\, line. In the upper right hand panel this correction was applied to the flux ratios for SDSS\,J0129. In the upper left hand panel the original flux ratios for SDSS\,J0129 are shown. A better agreement compared to other lines including the correction was found. The other three panels show the flux ratios, models with intersection for the other three systems (c) SDSS\,J1208 with corrected He\,{\sc i} 5015\,\AA\ line, d) SDSS\,J1642 with corrected He\,{\sc i} 5015\,\AA\ line, e) SDSS\,J1525 with corrected He\,{\sc i} 5015\,\AA\ line).}
    \label{fig:spot_fluxes}
 \end{minipage}
\end{figure*}
 

\subsection{Phase folded spectra and Doppler tomograms}\label{sec:doppler}
A bright spot, an accretion disc and sometimes a central spike feature are visible in the phase folded spectra and Doppler tomogram of an AM\,CVn system. The bright spot corresponds to the impact point where the accretion stream hits the disc. In this work the zero phase for the trailed spectra and the Doppler tomograms was set such that the bright spot is located in the upper left quarter of the Doppler tomogram and as such is arbitrary and not constrained by geometry such as could be the case in an eclipsing system.

Apart from the prominent central spike, SDSS\,J1208 also shows a second bright spot in most lines, best seen in He\,{\sc i} 7065\,\AA\ (Fig~\ref{fig:1208dopplermap}). The second bright spot is discussed in more detail in Sec. \ref{sec:2ndbrightspot}.
 
SDSS\,J0129 and SDSS\,J1525 show the bright spot and disc emission in all lines not contaminated by white dwarf absorption. Only some helium lines (4471\,\AA, 4685\,\AA, 5875\,\AA\ and 6678\,\AA) in SDSS\,J0129 show the central spike, whereas there is no indication for a central spike in SDSS\,J1525. SDSS\,J0129 and SDSS\,J1525 also show no indication for a second bright spot (Fig~\ref{fig:0129dopplermap}, \ref{fig:1525dopplermap}).


SDSS\,J1642 is different to the previous systems. It shows strong, narrow disc emission lines (Fig.~\ref{fig:1642dopplermap}). To make the bright spot visible, each individual spectrum was divided by the grand average spectrum. Individual lines do not show the bright spot very clearly. Only a combination of all strong helium lines (4471\,\AA, 4713\,\AA, 4921\,\AA, 5015\,\AA, 5875\,\AA, 6678\,\AA, 7065\,\AA, 7281\,\AA) reveals a clear S-wave with a fairly low amplitude ($\sim$400km\,s$^{-1}$). The strength of the S-wave is phase dependent (see Fig.~\ref{fig:1642dopplermap}). This is also seen in AM\,CVn itself \citep{nel01a, roe06a} and in SDSS\,J0804 \citep{roe09}.   

\subsubsection{Second bright spot in SDSS\,J1208}\label{sec:2ndbrightspot}
A double bright-spot feature is detected in the data of SDSS\,J1208 as has previously been found in e.g. SDSS\,J1240 \citep{roe05}, GP\,Com and V396\,Hya (Steeghs et al. in prep). These systems show the second bright spot feature at the same velocity as the first spot and, remarkably, all systems show a similar phase shift between the spots of about 120$^\circ$. 

For SDSS\,J1208, a phase-shift between the peaks of the spots of 125$^\circ$~$\pm$~23$^\circ$ was found using the method described in Sec.~\ref{sec:doppler_anal}. This result is consistent with the finding in \citet{roe05} for SDSS\,J1240. All three lines show that the integrated flux of the second bright spot has a similar intensity as the first bright spot.

\subsection{Mass ratio of SDSS\,J0129}
A superhump period of 37.9$\pm$0.2 min for SDSS\,J0129 was found by \citet{she10} during outburst. In combination with the orbital period identified here, this leads to a period excess ($\frac{P_{\rm sh}-P_{\rm orb}}{P_{\rm orb}}$) of $\epsilon$=0.0092$\pm$0.0054. The error for the excess was estimated from the accuracy on the periods and is dominated by the error on the superhump period. \citet{pat05} found an empirical relation ($\epsilon=0.18q+0.29q^2$) between the period excess and the mass ratio for a large number of hydrogen rich dwarf novae. Here, a mass ratio for SDSS\,J0129 of $q$=0.031$\pm$0.018 using this relation is obtained.   
\begin{figure} 
\begin{center}
\includegraphics[width=0.49\textwidth]{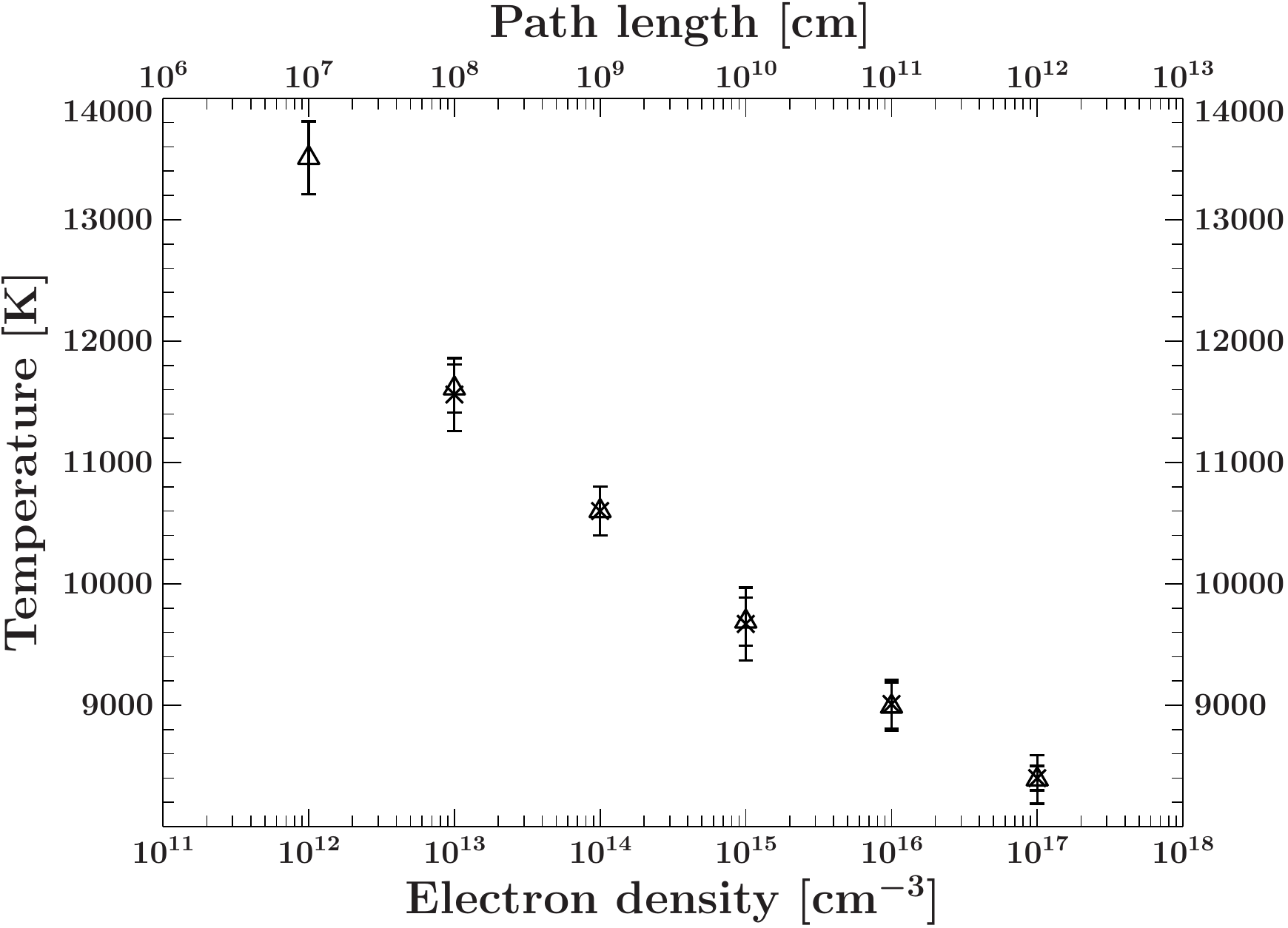}
\caption{Disc temperature estimation for SDSS\,J0129 with different assumptions for electron density and path length in the models. The X display the estimated temperatures with varied electron density and fixed path length (l=10$^9$cm). The triangles show the estimated temperatures with varied path length and fixed electron density ($n_{\rm e}$=10$^{14}$cm$^{-3}$).}
\label{fig:comparison}
\end{center}
\end{figure}

\subsection{Disc and bright spot temperature}\label{sec:temp}
To determine temperatures in the disc and bright spot the technique outlined in Sec.~\ref{sec:doppler_anal} was applied. For the disc emission, good agreement between models and observations within the error bars were found for SDSS\,J1208, SDSS\,J1642 and SDSS\,J0129 (see Fig.~\ref{fig:disc_do}). All three systems show similar temperatures in the disc: T$_{\rm disc}$=10\,800~$\pm$~400~K for SDSS\,J1208, T$_{\rm disc}$=10\,200~$\pm$~300~K for SDSS\,J1642 and T$_{\rm disc}$=10\,600~$\pm$~200~K for SDSS\,J0129 (see Tab.~\ref{tab:period}). The spectrum of SDSS\,J1525 shows a strong contribution from the accreting white dwarf in all lines bluewards of He\,{\sc i} 5015\,\AA\ (Fig.~\ref{gtcaver}). Therefore, He\,{\sc i} 5015\,\AA\ was excluded for the temperature estimate. SDSS\,J1525 is the only system that has no intersection for the flux ratio 6678/5875 in the disc region (see lower left panel in Fig.~\ref{fig:disc_do}) possibly due to contamination in the spectrum from the WD.

In contrast to the disc region, only poor agreement between models and observation was found for the bright spot region. Only SDSS\,J0129 shows good agreement between all flux ratios and the LTE model (see Fig.~\ref{fig:spot_fluxes}): T$_{\rm bright\,spot}$=27\,800~$\pm$~900~K. SDSS\,J1208 has only two intersection points with the models (6678/5015 and 5875/5015) which lie between 35\,000 and 40\,000 K (see middle left panel in Fig.~\ref{fig:spot_fluxes}). Therefore, no useful temperature estimation can be done. SDSS\,J1525 shows no intersection between models and observations at all. The flux ratios 7065/5015 and 5875/5015 from SDSS\,J1642 show discrepancy with the models. In particular, the first mentioned ratio shows a large discrepancy to the models (see middle right panel in Fig.~\ref{fig:spot_fluxes}).

\subsection{Column density variations}{\label{sec:column}}
To estimate systematic uncertainties on the temperatures with respect to electron density and the path length variations, two approaches were taken for SDSS\,J0129.

First, the temperatures were estimated by using models with a fixed path length at $l$=10$^9$cm and a variable electron density from $n_{\rm e}$=10$^{12}$cm$^{-3}$ to $n_{\rm e}$=10$^{17}$cm$^{-3}$. All show good agreement between models and observations except for the model with $n_{\rm e}$=10$^{12}$cm$^{-3}$. 

Second, the temperature was estimated by using models with a fixed electron density $n_{\rm e}$=10$^{14}$cm$^{-3}$ and a varied path length from $l$=10$^7$cm to $l$=10$^{12}$cm. A decreasing temperature with increasing path length or electron density is found (Fig.\ref{fig:comparison}). 

In both cases the column density, $n_{\rm e}$l, goes up, which increases the excitation of the helium lines as explained in \citet{mar91}. Helium lines further in the red have lower excitation energies than the helium lines in the blue. With increasing column density, the lines with lower excitation energy become stronger compared to the lines with higher excitation energies. To compensate for this, a lower temperature for a higher column density and a higher temperature for a lower column density is needed which can be seen in Fig.~\ref{fig:comparison}. This figure shows also that changing either the path length or the electron density leads to the same change in the estimated temperatures within the errorbars. But overall the temperature obtained is not very sensitive to a change in the path length or the electron density. A change in path length or electron density with a factor of 10\,000 only leads to temperature change of $\sim$3000~K.

\section{Conclusion \& Discussion}
\subsection{Metal absorption lines and spectral features}
All three known systems with periods between 50 and 60\,min (SDSS\,J1208, SDSS\,J1642 and SDSS\,J1552) display Mg absorption, whereas there is no evidence of this in the 65\,min period system V396\,Hya \citep{rui01} and 46\,min period system GP\,Com \citep{mar99}. Metal absorption lines are well known from single white dwarf spectra but for the temperature range where Mg\,{\sc i} is visible, also the Ca\,{\sc ii}~H~\&~K lines should be present. There is no evidence for calcium in either SDSS\,J1208 or SDSS\,J1642. 

Since the composition of the accretion disc and/or the upper layer of the primary white dwarf reflects the transferred material of the secondary, any abundance anomaly will reflect the current composition and/or the nucleosynthetic history of the secondary. This means that either magnesium was enhanced and/or calcium was depleted during the evolution of the system. Another possible explanation could be element selection, such as peeling off a sedimented secondary. Three different scenarios can account for the abundance anomaly:

1) If the secondary is a degenerate white dwarf, element sedimentation took place, such that heavy elements sunk down to the center of the white dwarf. An explanation for the strong magnesium lines and the lack of calcium could be that the accreted magnesium-rich material comes from a sedimented magnesium-rich layer of the secondary. Problematic for this picture is that e.g. SDSS\,J1208 also shows silicon which has a higher mass than magnesium and should have sunk down closer to the center of the WD than magnesium. An added complication is that it is also not clear if the secondary is fully convective at this stage of the evolution.

2) Element selection due to selective winds from the white dwarf could separate different elements. Selective winds driven by radiation pressure are found in early B-stars and blow away metals which have large cross sections \citep{hem03}. However, no strong winds in the temperature range are expected for accreting white dwarfs in AM\,CVn systems.

3) A third possible explanation could be differential gravitational settling of various elements. Different elements have different diffusion time scales. Depending on diffusion time scales compared to the mass accretion rate, abundances of heavy elements could increase at high accretion rates or decrease at low accretion rates compared to the diffusion timescales \citep{koe09}. \citet{bil06} pointed out that for systems with periods above 40\,min the nitrogen abundance will decrease in the atmosphere of the accretor because settling is faster than accretion. Therefore, systems with periods between 50-60\,min which already have low accretion rates ($\sim$10$^{-12}$M$_\odot$yr$^{-1}$) could be in the stage where settling is faster than accretion in the case of calcium, whereas for magnesium settling is slower than accretion. This would lead to a depletion of calcium and to an enhancement of magnesium in the atmosphere of the accreting WD. However, \citet{koe09} showed that the diffusion time scales for calcium are only slightly shorter than for magnesium.


In a comparison, good agreement is found for three of the four systems between the measured orbital period and expected spectral features based on other systems with a similar period. The only system which shows a discrepancy between  measured orbital period and expected spectral features is SDSS\,J1525, which has a spectrum similar to SDSS\,J0129 but a longer orbital period. One reason could be either a high mass accretor and/or a hot donor star because a high mass accretor is heated to a higher temperature due to compressional heating \citep{bil06}. 


\subsection{Mass ratio of SDSS\,J0129}
\citet{pat05} calibrated the relation between period excess and mass ratio ($\epsilon=0.18q+0.29q^2$) on a series of measurements of eclipsing Cataclysmic variables, where only KV\,UMa has a low ratio $q\le$0.05. For the extreme mass ratio regime this relation is uncertain.

In AM\,CVn systems, \citet{roe06a} found a large discrepancy for the mass ratio in AM\,CVn itself using either the empirical relation from \citet{pat05} ($q$=0.10) or obtaining the mass ratio from the projected velocity amplitude and the phase of the accreting white dwarf ($q\geq$0.18). \citet{cop11} found agreement within the uncertainties in the eclipsing AM\,CVn system SDSS\,J092638.71+362402.4 between the mass ratio obtained from the phase of the white dwarf eclipse and the mass ratio obtained from the relation from \citet{pat05}. At this stage too few mass ratios are known to confirm or reject this relation for AM\,CVn systems. 

\subsection{Disc and bright spot temperatures}
The single-slab LTE models with uniform density and temperature are a clear simplification of the real environment in an accretion disc. Nevertheless, the good agreement between models and observations for three out of four systems shows that the flux in the helium disc lines are well fitted with the simplified models. This leads to the assumption that the line emission possibly occurs in a small, nearly isothermal, region in the upper layers of the disc. Under the assumption of a fixed path length of $l$=10$^9$cm and an electron density of 10$^{14}$cm$^{-1}$, similar temperatures as \citet{mar91} in GP\,Com are found. For the disc temperatures between 15\,000~K and 25\,000~K the flux ratio of the helium lines used here is rather insensitive to temperature variations. Therefore, this method can not be used for discs with temperatures within this range. Contrary to the disc region, the flux ratios from the bright spot region show good agreement between models and observations in only one system. This shows that the LTE assumption with uniform density and temperature is not completely valid in the bright spot region. 

\section*{Acknowledgements}
TK acknowledges support by the Netherlands Research School for Astronomy (NOVA). TRM and DS acknowledge the support from the Science and Technology Facilities Council (STFC) during the course of this work. GN acknowledges an NWO-VIDI grant.

Based on observations made with the Gran Telescopio Canarias (GTC), installed in the Spanish Observatorio del Roque de los Muchachos of the Instituto de Astrof´ısica de
Canarias, in the island of La Palma. Some results presented in this paper are based on observations made with the William Herschel Telescope operated on the island of La Palma by the Isaac Newton Group in the Spanish Observatorio del Roque de los Muchachos of the Institutio de Astrofisica de Canarias. Some of the data presented herein were obtained at the Palomar Observatory, California Institute of Technology.
\nocite{ste12}





\bibliography{refs}{}
\bibliographystyle{mn2e}


\label{lastpage}

\end{document}